\documentclass[prb,twocolumn,groupedaddress,shownopacs,amssymb]{revtex4-2}
\usepackage{graphicx,latexsym}
\usepackage{amsmath,amssymb,amsfonts}
\usepackage{bm}
\usepackage{braket}
\usepackage{mathrsfs}

\newcommand{\up}{\uparrow}
\newcommand{\down}{\downarrow}
\newcommand{\m}{\mathcal}
\newcommand{\ep}{\varepsilon}
\begin{document}
\title{Time-local nonequilibrium Green's function method for real-time dynamics in quantum systems coupled to superconducting leads}

\author{Taira Kawamura} 
\affiliation{Department of Physics, College of Science and Technology, Nihon University, Tokyo 101-8308, Japan}

\date{\today}

\begin{abstract}
We develop a time-local nonequilibrium Green's function formulation for real-time dynamics in quantum systems coupled to superconducting leads. The superconducting lead self-energy is a strongly frequency-dependent matrix in Nambu space, giving rise to nonlocal memory kernels in the time domain. This makes direct propagation of the Kadanoff–Baym (KB) equations computationally demanding. To overcome this difficulty, we extend the auxiliary-mode expansion, originally developed for normal-metal leads, to Nambu-space self-energies. This allows us to decompose the superconducting lead self-energy into a finite number of exponential modes and to transform the KB equations with memory integrals into a closed set of ordinary differential equations. The resulting time-local equations enable efficient real-time simulations under general time-dependent bias voltages, superconducting phases, and one-body Hamiltonians of the central system, while retaining the memory effects induced by superconducting leads. As an application, we analyze voltage-quench dynamics in a superconductor–quantum-dot–superconductor junction and show that, after a dc bias is suddenly applied, the system evolves through a transient regime and relaxes to an ac Josephson periodic steady state. The resulting periodic steady-state current agrees with the Floquet Green's function solution, validating the present real-time formulation.
\end{abstract}

\maketitle
\section{Introduction}
\label{sec.Intro}

Hybrid structures combining superconductors with semiconductor nanowires, quantum dots, or two-dimensional electron systems provide a versatile platform for studying coherent transport phenomena associated with the Josephson effect and Andreev reflection~\cite{Josephson1962, TinkhamBook, HeikkilaBook, Lambert1998, Golubov2004, Rodero2011}. In these systems, a wide range of nonequilibrium phenomena has been studied, including the ac Josephson effect~\cite{Giaever1965, Holst1994, Billangeon2007, Woerkom2017, Laroche2019, Haller2023}, Shapiro responses~\cite{Shapiro1963, Shapiro1964, Ueda2020, Dartiailh2021, Iorio2023, Wu2025}, microwave spectroscopy of Andreev bound states~\cite{Bretheau2013, Bretheau2013_PRX, Janvier2015, Van2017, Hays2018, Park2022}, and photon-assisted subgap transport~\cite{Chauvin2006, Peters2020, Carrad2022, Haxell2023}. A theoretical understanding of these nonequilibrium phenomena requires a microscopic framework for describing the dynamics of superconducting junctions under general time-dependent driving.

The nonequilibrium Green's function (NEGF) method provides a powerful microscopic framework for describing nonequilibrium dynamics in open quantum systems~\cite{Datta1995, Haug2008, Stefanucci2013}. Indeed, time-dependent transport has been extensively studied within the NEGF framework for systems coupled to normal-metal leads~\cite{Jauho1994, Zhu2005, Maciejko2006, Odashima2017, Ridley2022}. Extending this framework to superconducting leads, however, is substantially more challenging. The self-energy of a superconducting lead is a strongly frequency-dependent Nambu matrix, reflecting both the gap structure in the quasiparticle density of states and anomalous correlations that mix particle and hole degrees of freedom. In the time domain, this frequency dependence produces a nonlocal memory kernel in the Kadanoff–Baym (KB) equations, which govern the real-time evolution of Green's functions in the NEGF formalism. Direct propagation of the resulting KB equations is thus computationally demanding in systems coupled to superconducting leads.

A useful strategy for handling such nonlocal memory kernels is the auxiliary-mode expansion, which was originally developed for systems coupled to normal-metal leads~\cite{Croy2009, Popescu2016, Pavlyukh2025, Croy2011, Croy2012, Tuovinen2023, Kawamura2024, Kawamura2025}. In this approach, the Fermi distribution function and the lead density of states are approximated by pole expansions, allowing the lead self-energy to be decomposed into a finite number of exponential modes in the time domain. The memory integrals in the KB equations are then converted into time-local equations of motion (EOMs) for auxiliary modes, enabling efficient finite-temperature simulations of time-dependent transport. Existing auxiliary-mode approaches, however, have been formulated mainly for normal-metal leads and typically rely on Lorentzian decompositions of the lead density of states~\cite{Croy2009, Popescu2016, Pavlyukh2025}. As a result, they cannot be directly applied to superconducting leads, whose self-energies are strongly frequency-dependent matrices in Nambu space. Extending the auxiliary-mode approach to Nambu-space self-energies and general lead spectral functions is therefore a central step toward time-dependent transport theory with superconducting leads.

In this work, we extend the auxiliary-mode approach to superconducting leads and derive time-local EOMs for quantum systems coupled to them. Using rational approximations based on the adaptive Antoulas–Anderson algorithm~\cite{Nakatsukasa2018}, we express general Nambu-space lead self-energies as finite sums of exponential modes in the time domain. This transformation converts the time-nonlocal memory integrals in the KB equations into a closed set of time-local EOMs involving auxiliary functions. The resulting method enables efficient real-time simulations under general time-dependent driving, including time-dependent lead voltages, superconducting phases, and central-system Hamiltonians, while retaining the non-Markovian memory effects induced by superconducting leads.

As a concrete application, we study voltage-quench dynamics in a superconducting junction where a single-level quantum dot is coupled to two superconducting leads. After a dc bias is suddenly applied across the junction, the system relaxes in the long-time limit to a periodic steady state characterized by the ac Josephson frequency. We also show that subharmonic components can emerge in the transient regime and gradually decay as the system approaches the ac Josephson steady state. The long-time periodic state is benchmarked against the corresponding solution obtained from the Floquet Green's function method~\cite{Sun2002}.

Before closing the introduction, we briefly discuss how the present approach is related to other microscopic methods for nonequilibrium dynamics in superconducting junctions. For Josephson junctions under a constant voltage bias, the long-time state is an ac Josephson periodic steady state, which can be efficiently described by the Floquet Green's function method~\cite{Sun2002}. Transient dynamics can also be treated by using the final stationary or periodic steady state as a reference state~\cite{Cheng2024}. These approaches, however, are not directly applicable when the final state is not known in advance or when the driving is not strictly periodic, as in pulse, ramp, or noisy-bias protocols. Time-dependent subgap dynamics has also been studied analytically by solving Heisenberg equations of motion, although in the infinite-gap limit of the superconducting leads~\cite{Taranko2019}. Real-time propagation schemes based on the equivalence between the NEGF formalism and the time-dependent Bogoliubov–de Gennes equation have also been developed for junctions with semi-infinite superconducting leads and applied to nonequilibrium Josephson dynamics~\cite{Stefanucci2010}. The present approach is complementary to these methods: it retains the transparent NEGF formalism while recasting the memory-kernel problem into a computationally tractable time-local form.

The remainder of this paper is organized as follows. In Sec.~\ref{sec.Formalism}, we introduce a general model of a quantum system coupled to multiple superconducting leads and derive time-local EOMs based on the NEGF method and the auxiliary-mode expansion. In Sec.~\ref{sec.application}, we apply the time-local EOMs to voltage-quench dynamics in a superconductor–quantum-dot–superconductor junction. Throughout this paper, we set $\hbar=k_{\rm B}=1$ and take the volume of each superconducting lead to be unity.

\section{Formalism}
\label{sec.Formalism}

\subsection{Hamiltonian}
\label{sec.Hamiltonian}

\begin{figure}[t]
\centering
\includegraphics[width=\linewidth]{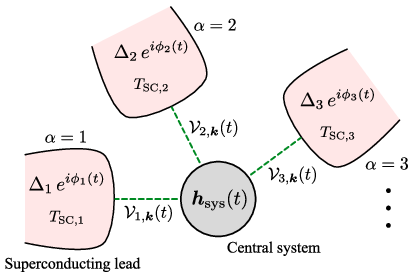}
\caption{
Schematic illustration of the general setup considered in this work. A central quantum system described by a one-body Hamiltonian $\bm{h}_{\rm sys}(t)$ is coupled to multiple macroscopic superconducting leads labeled by $\alpha=1,\ldots,N_{\rm lead}$. Each superconducting lead is assumed to remain in thermal equilibrium at temperature $T_{{\rm SC},\alpha}$ and is characterized by a gap amplitude $\Delta_\alpha$ and a superconducting phase $\phi_\alpha(t)$. The phase $\phi_\alpha(t)$ may be a general function of time and is incorporated into the tunneling matrix $\mathcal{V}_{\alpha,\bm{k}}(t)$ between the central system and lead $\alpha$.
}
\label{Fig_model} 
\end{figure}

We consider a general quantum system coupled to multiple macroscopic superconducting leads, as schematically illustrated in Fig.~\ref{Fig_model}. The total Hamiltonian is
\begin{equation}
H(t)
=
H_{\rm sys}(t)
+
\sum_{\alpha=1}^{N_{\rm lead}} H_{{\rm SC},\alpha}
+
\sum_{\alpha=1}^{N_{\rm lead}} H_{{\rm T},\alpha}(t),
\end{equation}
where $H_{\rm sys}(t)$ describes the central system, $H_{{\rm SC},\alpha}$ is the Hamiltonian of superconducting lead $\alpha$, and $H_{{\rm T},\alpha}(t)$ describes the tunneling between the central system and lead $\alpha$. Here, $\alpha=1,\ldots,N_{\rm lead}$ labels the superconducting leads.

The Hamiltonian of the central system is written in the Nambu representation as
\begin{equation}
H_{\rm sys}(t)=
\Psi^\dagger\, 
\bm{h}_{\rm sys}(t)\,
\Psi.
\label{eq.Hsys}
\end{equation}
Here,
\begin{equation}
\Psi
=
\begin{pmatrix}
\psi_1 &
\psi_2 &
\cdots &
\psi_N
\end{pmatrix}^{\rm T},
\quad
\psi_j
=
\begin{pmatrix}
c_{j,\uparrow} &
c^\dagger_{j,\downarrow}
\end{pmatrix}^{\rm T},
\end{equation}
is the $2N$-component Nambu field in the orbital $\otimes$ Nambu space, where $N$ is the number of single-particle orbitals in the central system. In Eq.~\eqref{eq.Hsys}, $\bm{h}_{\rm sys}(t)$ is a general $2N\times 2N$ single-particle Hamiltonian. We assume that time-dependent fields in the central system are absent for $t<0$. For $t>0$, $\bm{h}_{\rm sys}(t)$ may include general time-dependent one-body terms, such as externally applied fields and mean-field order parameters.

Each superconducting lead is described by the mean-field BCS Hamiltonian
\begin{equation}
H_{{\rm SC},\alpha}=
\sum_{\bm k}
\Phi^\dagger_{\alpha,\bm k}\,
h_{\alpha,\bm k}\,
\Phi_{\alpha,\bm k},
\label{eq:Hsc}
\end{equation}
where
\begin{equation}
\Phi_{\alpha,\bm k}
=
\begin{pmatrix}
a^\alpha_{\bm k,\uparrow}\\[4pt]
a^{\alpha\dagger}_{-\bm k,\downarrow}
\end{pmatrix}
\end{equation}
is the two-component Nambu spinor in lead $\alpha$. The corresponding single-particle Hamiltonian is
\begin{equation}
h_{\alpha,\bm k}
=
\begin{pmatrix}
\xi_{\alpha,\bm k} & -\Delta_\alpha\\[4pt]
-\Delta_\alpha & -\xi_{\alpha,\bm k}
\end{pmatrix},
\end{equation}
where $\xi_{\alpha,\bm k}$ is the single-particle energy and $\Delta_\alpha$ is the superconducting gap amplitude. We choose a gauge in which $\Delta_\alpha$ is real in each lead~\cite{Sun2002, Sun2000, Sun2000_2}. The superconducting phase and the bias voltage applied to lead $\alpha$ are incorporated into the time-dependent tunneling Hamiltonian $H_{{\rm T},\alpha}(t)$ introduced below. The gap amplitude $\Delta_\alpha$ may depend on the lead index $\alpha$ but is assumed to be time independent. The quasiparticle distribution in lead $\alpha$ is taken to be the Fermi–Dirac distribution at temperature $T_{{\rm SC},\alpha}$,
\begin{equation}
f_\alpha(\omega) = 
\frac{1}{e^{\omega/T_{{\rm SC},\alpha}} +1}.
\label{eq.Fermi}
\end{equation}

The tunneling Hamiltonian between the central system and lead $\alpha$ is
\begin{equation}
H_{{\rm T},\alpha}(t)
=
\sum_{\bm k}
\left[
\Psi^\dagger\,
\m{V}_{\alpha,\bm k}(t)\,
\Phi_{\alpha,\bm k}
+
{\rm H.c.}
\right].
\label{eq.HT}
\end{equation}
Here, $\m{V}_{\alpha,\bm k}(t)\in \mathbb C^{2N\times 2}$ is the tunneling matrix connecting the orbital $\otimes$ Nambu space of the central system to the Nambu space of superconducting lead $\alpha$. In the gauge adopted above, the superconducting phase $\phi_\alpha(t)$ of lead $\alpha$ is incorporated into the tunneling matrix as
\begin{equation}
\m{V}_{\alpha,\bm{k}}(t) =
\m{T}_{\alpha,\bm{k}} \ket{e_\alpha} \otimes U_\alpha(t)\, \tau_z,
\label{eq.V}
\end{equation}
with
\begin{equation}
U_\alpha(t)
=
\begin{pmatrix}
e^{-i\phi_{\alpha}(t)/2} & 0 \\
0 & e^{i\phi_{\alpha}(t)/2}
\end{pmatrix}.
\label{eq.U}
\end{equation}
Here, $\tau_z$ is the Pauli matrix in Nambu space, and $\m{T}_{\alpha,\bm k}$ is the tunneling amplitude between the central system and lead $\alpha$. The vector $\ket{e_\alpha}\in\mathbb R^{N\times 1}$ specifies the orbital channel coupled to lead $\alpha$. For example, for a one-dimensional tight-binding chain connected to left ($\alpha=1$) and right ($\alpha=2$) superconducting leads at its two ends, one may choose
\begin{equation}
\ket{e_{\alpha=1}}=\ket{1},
\qquad
\ket{e_{\alpha=2}}=\ket{N},
\end{equation}
where $\ket{j}$ denotes the orbital localized at site $j$.

The superconducting phase $\phi_\alpha(t)$ in Eq.~\eqref{eq.U} is treated as a given function of time. For $t\leq 0$, all superconducting phases are assumed to be fixed at constant values, $\phi_\alpha(t)=\phi_\alpha(0)$. The system is then in an initial equilibrium state, although a dc Josephson current may flow when finite phase differences are present between superconducting leads. For $t>0$, the phases $\phi_\alpha(t)$ may become time dependent, and the resulting time dependence of the tunneling Hamiltonian can drive the system out of equilibrium. Such time-dependent phases may be generated, for example, by a time-dependent magnetic flux or by voltage-biased superconducting leads. When a bias voltage $V_\alpha(t)$ is applied to lead $\alpha$ for $t>0$, the phase obeys~\cite{Josephson1962, TinkhamBook}
\begin{equation}
\frac{d \phi_\alpha(t)}{dt}=
2eV_\alpha(t),
\label{eq.Josephson.relation}
\end{equation}
which gives
\begin{equation}
\phi_\alpha(t)
=
\phi_\alpha(0)
+
2e\int_0^t dt_1\, V_\alpha(t_1).
\label{eq.phi.V}
\end{equation}

\begin{table*}[t]
\setlength{\belowcaptionskip}{8pt}
\caption{
Summary of the main quantities in the time-local formulation. Detailed definitions are given in the equations indicated in the last column.
}
\label{tab:notation_overview}
\setlength{\tabcolsep}{8pt}
\begin{tabular}{lll}
\hline
Symbol & Meaning & Definition \\
\hline
$\bm{G}^<(t)$
& Equal-time lesser Green's function
& Eq.~(\ref{eq.def.equal.GL}) \\[2pt]

$\bm{\Pi}_{K}(t)$
& Current matrix for auxiliary mode $K$
& Eq.~(\ref{eq.Pi.AME}) \\[2pt]

$\bm{\Omega}_{K, K'}(t)$
& Auxiliary matrix coupling modes $K$ and $K'$
& Eq.~(\ref{eq.Omega.KK}) \\[2pt]

$\chi_{K}$
& Pole in the auxiliary-mode expansion
& Eq.~(\ref{eq.def.chi}) \\[2pt]

$R_K^{\lessgtr}$
& Residue of the lesser/greater kernel
& Eqs.~\eqref{eq.def.Rl} and \eqref{eq.def.Rg} \\[2pt]

$\bm{P}_K(t)$
& Orbital--Nambu coupling matrix
& Eq.~(\ref{eq.def.P}) \\[2pt]

$\bm{\Phi}_K(t)$
& Superconducting phase-velocity matrix
& Eq.~(\ref{eq.def.Phi.mat}) \\

\hline
\end{tabular}
\end{table*}

\subsection{NEGF of the central system}

To describe nonequilibrium transport in the quantum device, we introduce the lesser and greater Green's functions of the central system. In the orbital $\otimes$ Nambu representation, they are $2N\times 2N$ matrices defined by~\cite{Datta1995, Haug2008, Stefanucci2013}
\begin{align}
&
\bm{G}^<(t,t') =i \braket{\Psi^\dagger(t')\Psi(t)}
,\\[4pt]
&
\bm{G}^>(t,t') = -i \braket{\Psi(t) \Psi^\dagger(t')}.
\end{align}
In the following formulation, the equal-time lesser Green's function,
\begin{equation}
\bm{G}^<(t)
\equiv
\bm{G}^<(t,t),
\label{eq.def.equal.GL}
\end{equation}
plays a central role. It contains the one-body density matrix of the central region and directly determines local physical quantities. For example, its diagonal particle components give the electron populations of the corresponding orbitals, while its off-diagonal Nambu components describe anomalous correlations when superconducting pairing is induced in the central system.

As derived in Appendix~\ref{sec.app.KB}, the time evolution of $\bm{G}^<(t)$ is governed by the equal-time KB equation~\cite{Kawamura2024, Kawamura2025}
\begin{equation}
i \frac{d}{dt} \bm{G}^<(t)=
\big[\bm{h}_{\rm sys}(t),\, \bm{G}^<(t)\big]_-
-
\sum_{\alpha=1}^{N_{\rm lead}}
\big[
\bm{\Pi}_\alpha(t) +
\bm{\Pi}^\dagger_\alpha(t)
\big],
\label{eq.KB}
\end{equation}
where $[A,B]_-=AB-BA$. The first term on the right-hand side describes coherent evolution within the central system, whereas the second term represents the effect of the superconducting leads. The matrix $\bm{\Pi}_\alpha(t)$ is defined as
\begin{align}
\bm{\Pi}_\alpha(t) 
&=
\int_{-\infty}^t dt_1\,
\big[
\bm{G}^>(t,t_1)\, \bm{\Sigma}^<_\alpha(t_1,t)
\notag\\
&\hspace{3cm}-
\bm{G}^<(t,t_1)\, \bm{\Sigma}^>_\alpha(t_1,t)
\big].
\label{eq.Pi}	
\end{align}
Here, $\bm{\Sigma}^{\lessgtr}_\alpha(t,t')$ are the lesser and greater self-energies due to the coupling to lead $\alpha$. Their explicit forms are given in Sec.~\ref{sec.self}.

The physical meaning of $\bm{\Pi}_\alpha(t)$ is clarified by its relation to the electric current. The current $J_\alpha(t)$ flowing from lead $\alpha$ into the central system is given by~\cite{Datta1995, Haug2008, Stefanucci2013}
\begin{align}
J_\alpha(t)
&=
2e\, {\rm Re} 
\int_{-\infty}^t dt_1\,
{\rm Tr}\Big[
\big( I_N \otimes \tau_z\big)
\notag\\
&\qquad \times
\big[
\bm{G}^>(t, t_1) \bm{\Sigma}^<_{\alpha}(t_1,t) 
-
\bm{G}^<(t, t_1) \bm{\Sigma}^>_{\alpha}(t_1,t)
\big]
\Big]
\notag\\[4pt]
&=
2e\, {\rm Re}\, 
{\rm Tr}\big[
\big( I_N \otimes \tau_z\big)\,
\bm{\Pi}_\alpha(t)\big],
\label{eq.current.J}
\end{align}
where $I_N$ is the $N\times N$ identity matrix in orbital space, and the trace is taken over the orbital and Nambu degrees of freedom. Thus, $\bm{\Pi}_\alpha(t)$ can be regarded as a current matrix associated with lead $\alpha$.

Equation~\eqref{eq.KB} has the form of an integro-differential equation. Although it governs the equal-time Green's function $\bm{G}^<(t)$, the lead contribution $\bm{\Pi}_\alpha(t)$ in Eq.~\eqref{eq.Pi} contains an integral over the entire past history through the two-time Green's functions $\bm{G}^{\lessgtr}(t,t_1)$ and self-energies $\bm{\Sigma}^{\lessgtr}_\alpha(t_1,t)$. This non-Markovian structure is essential for superconducting leads, because the lead self-energies encode the nontrivial energy dependence of the superconducting density of states and the quasiparticle distribution functions. At the same time, it makes direct numerical propagation highly demanding, since one has to store and propagate two-time quantities. In the following subsections, we show how this non-Markovian equation can be rewritten as a closed set of ordinary differential equations by introducing auxiliary dynamical variables.

\subsection{Time-local equations of motion}

Before giving the detailed derivation, we first present the main result of this work: a closed set of time-local equations of motion (EOMs). This provides an overview of the proposed propagation scheme and clarifies which quantities must be propagated in time. Some quantities appearing in these equations are introduced in the subsequent subsections. For the reader's convenience, Table~\ref{tab:notation_overview} summarizes the main symbols and indicates where they are defined.

The non-Markovian Kadanoff–Baym equation~\eqref{eq.KB} can be rewritten as the following coupled ordinary differential equations:
\begin{widetext}
\begin{subequations}
\begin{align}
&
i \frac{d}{dt} \bm{G}^<(t) =
\big[\bm{h}_{\rm sys}(t),\, \bm{G}^<(t)\big]_-
- 
\sum_{K} \big[\bm{\Pi}_{K}(t) + \bm{\Pi}^\dagger_{K}(t) \big]
\label{eq.EOM.GL}
,\\
&
i\frac{d}{dt}\, \bm{\Pi}_{K}(t)
=
-R^<_{K}\,
\bm{P}_{K}(t)
+
i\big[R^>_{K} - R^<_{K}\big]
\bm{G}^<(t)\, \bm{P}_{K}(t)
+
\big[\bm{h}_{\rm sys}(t) - \chi_{K}\big]
\bm{\Pi}_{K}(t)
-
\bm{\Pi}_{K}(t)\,
\bm{\Phi}_K(t)
+
\sum_{K'} \bm{\Omega}_{K,K'}(t)
\label{eq.EOM.Pi}
,\\
&
i \frac{\partial}{\partial t}\,
\bm{\Omega}_{K, K'}(t)=
-i
\big[R^>_{K} -R^<_{K}\big]\, 
\bm{\Pi}^\dagger_{K'}(t)\,
\bm{P}_{K}(t)
+
i
\big[\big(R^>_{K'}\big)^* -\big(R^<_{K'}\big)^*\big]
\bm{P}_{K'}(t)\,
\bm{\Pi}_{K}(t)
\notag\\[4pt]
&\hspace{4.5cm}+
\big[
\chi^*_{K'} - \chi_{K}
\big]\,
\bm{\Omega}_{K,K'}(t)
+
\bm{\Phi}_{K'}(t)\,
\bm{\Omega}_{K, K'}(t)
-
\bm{\Omega}_{K, K'}(t)\,
\bm{\Phi}_K(t).
\label{eq.EOM.Omega}
\end{align}
\end{subequations}
\end{widetext}
Here, $K=\{\alpha,\nu,l\}$ is a composite index labeling the lead $\alpha$ $(=1,\ldots,N_{\rm lead})$, the Nambu branch $\nu$ $(=\pm)$, and the pole $l$ $(=1,\ldots,N^\alpha_{\rm pole})$ appearing in the auxiliary-mode expansion introduced below. The total number of auxiliary modes is
\begin{equation}
N_{\rm tot}=
2\sum_{\alpha=1}^{N_{\rm lead}} N^\alpha_{\rm pole}.
\end{equation}
For notational simplicity, quantities that do not depend on all components of the composite index $K$ are also labeled by $K$. In particular, we write
\begin{equation}
\bm{P}_K(t)\equiv \bm{P}_{\alpha,\nu}(t),
\qquad
\bm{\Phi}_K(t)\equiv \bm{\Phi}_\alpha(t).
\end{equation}

Equations~\eqref{eq.EOM.GL}–\eqref{eq.EOM.Omega} are the central working equations of the present formalism. The time-dependent matrices $\bm{P}_K(t)$ and $\bm{\Phi}_K(t)$ are given functions of time, determined by the superconducting phases $\phi_\alpha(t)$. The quantities to be propagated are the equal-time lesser Green's function $\bm{G}^<(t)$, the current matrices $\bm{\Pi}_K(t)$, and the auxiliary matrices $\bm{\Omega}_{K,K'}(t)$. The original memory integral in Eq.~\eqref{eq.Pi} is thereby replaced by time-local auxiliary variables, so that the time evolution can be performed without propagating and storing the full two-time Green's functions. This reformulation substantially reduces the computational and memory cost. The initial conditions at $t=0$ for these coupled ordinary differential equations are discussed in Sec.~\ref{sec.Initial}.

\subsection{Derivation of the self-energy $\bm{\Sigma}^\lessgtr_\alpha(t,t')$}
\label{sec.self}

We first derive the lesser and greater self-energies $\bm{\Sigma}^{\lessgtr}_\alpha(t,t')$ originating from the coupling to superconducting lead $\alpha$. 
Since the superconducting leads are described by the quadratic BCS mean-field Hamiltonian in Eq.~\eqref{eq:Hsc} and the tunneling Hamiltonian in Eq.~\eqref{eq.HT} is bilinear, the lead degrees of freedom can be integrated out exactly. The resulting embedding self-energy due to the coupling to lead $\alpha$ is given by~\cite{Rodero2011}
\begin{equation}
\bm{\Sigma}^\lessgtr_\alpha(t, t')
=
\sum_{\bm{k}}\,
\m{V}_{\alpha,\bm{k}}(t)\,
D_{\alpha}^\lessgtr(\bm{k},t -t')\,
\m{V}^\dagger_{\alpha,\bm{k}}(t').
\label{eq.self}
\end{equation}
Here, $D_{\alpha}^{\lessgtr}$ is the $2\times 2$ Nambu Green's function of lead $\alpha$. Since the superconducting lead is assumed to remain in thermal equilibrium, $D_{\alpha}^{\lessgtr}$ depends only on the relative time. Using the fluctuation–dissipation relation~\cite{Datta1995, Haug2008, Stefanucci2013}, the lesser and greater Green's functions of the superconducting lead are expressed as
\begin{subequations}
\begin{align}
&
D^<_{\alpha}(\bm{k}, t-t')=
-\int_{-\infty}^\infty \frac{d\omega}{2\pi}\,
e^{-i\omega(t-t')}f_\alpha(\omega)
\notag\\
&\hspace{3cm}\times
\big[
D^{\rm R}_\alpha(\bm{k},\omega) -
D^{\rm A}_\alpha(\bm{k},\omega)
\big]
,\\[8pt]
&
D^>_{\alpha}(\bm{k}, t-t')=
\int_{-\infty}^\infty \frac{d\omega}{2\pi}\,
e^{-i\omega(t-t')}
f_\alpha(-\omega)
\notag\\
&\hspace{3cm}\times
\big[
D^{\rm R}_\alpha(\bm{k},\omega) -
D^{\rm A}_\alpha(\bm{k},\omega)
\big],
\end{align}
\end{subequations}
where $f_\alpha(\omega)$ is the Fermi–Dirac distribution function defined in Eq.~\eqref{eq.Fermi}. The retarded and advanced Green's functions are given by~\cite{Rodero2011}
\begin{subequations}
\begin{align}
D^{\rm R}_\alpha(\bm{k}, \omega)
&=
\frac{1}{[\omega +i\Gamma_\alpha]^2 -\xi^2_{\alpha,\bm{k}} -\Delta_\alpha^2}
\notag\\
&\quad\times
\begin{pmatrix}
\omega +i\Gamma_\alpha +\xi_{\alpha,\bm{k}} & 
-\Delta_\alpha \\
-\Delta_\alpha & 
\omega +i\Gamma_\alpha -\xi_{\alpha,\bm{k}}
\end{pmatrix}
\label{eq.DR}
,\\
D^{\rm A}_\alpha(\bm{k}, \omega)
&=
\big[D^{\rm R}_\alpha(\bm{k}, \omega)\big]^\dagger.
\end{align}
\end{subequations}
Here, $\Gamma_\alpha$ is the Dynes broadening parameter~\cite{Dynes1978}, which phenomenologically accounts for finite quasiparticle lifetimes in lead $\alpha$. This phenomenological description is widely used for real superconductors, in which a nonzero $\Gamma_\alpha$ arises naturally from inelastic scattering processes~\cite{Dynes1978}. Moreover, as discussed in Sec.~\ref{sec.AME}, a nonzero $\Gamma_\alpha$ is essential for the auxiliary-mode expansion developed below. Throughout this work, we consider the regime $\Gamma_\alpha\ll\Delta_\alpha$, in which the superconducting gap remains well defined.

We perform the momentum summation in Eq.~\eqref{eq.self} by converting it into an integral over the single-particle energy $\xi_{\alpha}$:
\begin{equation}
\sum_{\bm{k}} \to N_\alpha(0) \int d\xi_\alpha\, W_{\rm c}(\xi_\alpha).
\label{eq.ksum.cut}
\end{equation}
Here, $N_\alpha(0)$ is the normal-state density of states at the Fermi level in lead $\alpha$, and $W_{\rm c}(\xi_\alpha)$ is a cutoff function that specifies the effective bandwidth of the superconducting lead. In many transport calculations, one takes a flat normal-state density of states and assumes the wide-band limit, $W_{\rm c}(\xi_\alpha)=1$~\cite{Datta1995, Haug2008, Stefanucci2013}. In the present formulation, however, it is useful to keep a finite bandwidth, as will become clear in the auxiliary-mode expansion discussed in Sec.\ref{sec.AME}. We thus introduce a smooth cutoff function and use, in this work, the super-Gaussian form
\begin{equation}
W_{\rm c}(\xi_\alpha) = 
\exp\left(
-\frac{\xi^{2n}_\alpha}{D^{2n}}
\right),
\label{eq.cutoff}
\end{equation}
where $D$ characterizes the bandwidth and $n$ controls the sharpness of the cutoff. The numerical results of the time-evolution simulations are insensitive to the specific choice of $W_{\rm c}$, provided that the cutoff scale $D$ is sufficiently larger than the relevant low-energy scales.

Assuming that the tunneling amplitude is momentum independent, $\m{T}_{\alpha,\bm{k}}=\m{T}_{\alpha}$, and replacing the momentum summation in Eq.~\eqref{eq.self} with an integral over $\xi_\alpha$ according to Eq.~\eqref{eq.ksum.cut}, we can write the self-energy as
\begin{equation}
\bm{\Sigma}^{\lessgtr}_{\alpha}(t,t')
=
-i\gamma_{\alpha}\,
\ket{e_{\alpha}}\bra{e_{\alpha}}
\otimes
U_{\alpha}(t)
\mathscr{D}^{\lessgtr}_{\alpha}(t-t')
U^{\dagger}_{\alpha}(t').
\label{eq.self.xi}
\end{equation}
Here, we have introduced
\begin{equation}
\gamma_\alpha \equiv \pi N_\alpha(0)|\m{T}_\alpha|^2,
\label{eq.gamma.def}
\end{equation}
and the $\xi_\alpha$-integrated Nambu Green's function
\begin{equation}
\mathscr{D}^\lessgtr_{\alpha}(t-t')
=
\frac{i}{\pi}
\int_{-\infty}^{\infty}d\xi_{\alpha}\,
W_{\rm c}(\xi_{\alpha})\,
\tau_z
D^\lessgtr_{\alpha}
\bigl(\xi_{\alpha},t-t'\bigr)
\tau_z.
\label{eq.D.integrated}
\end{equation}
We note that the two factors of $\tau_z$ arising from the tunneling matrices in Eq.~\eqref{eq.V} are included in the definition of $\mathscr{D}^{\lessgtr}_{\alpha}(t-t')$.

The retarded counterpart of Eq.~\eqref{eq.D.integrated} can be evaluated using Eq.~\eqref{eq.DR} as
\begin{align}
\mathscr{D}^{\rm R}_{\alpha}(\omega)
&=
\frac{i}{\pi}
\int_{-\infty}^{\infty}d\xi_{\alpha}\,
\frac{W_{\rm c}(\xi_\alpha)}{[\omega +i\Gamma_\alpha]^2 -\xi^2_{\alpha} -\Delta_\alpha^2}
\notag\\
&\qquad\times
\begin{pmatrix}
\omega +i\Gamma_\alpha +\xi_{\alpha} & 
\Delta_\alpha \\
\Delta_\alpha & 
\omega +i\Gamma_\alpha -\xi_{\alpha}
\end{pmatrix}
\notag\\
&=
\beta_\alpha(\omega)\,
\begin{pmatrix}
1 & \dfrac{\Delta_\alpha}{\omega +i\Gamma_\alpha}  \\
\dfrac{\Delta_\alpha}{\omega +i\Gamma_\alpha} & 1
\end{pmatrix},
\label{eq.qc.DR}
\end{align}
where
\begin{equation}
\beta_\alpha(\omega) =
\frac{i[\omega +i\Gamma_\alpha]}{\pi}
\int_{-\infty}^\infty d\xi_\alpha\,
\frac{W_{\rm c}(\xi_\alpha)}{[\omega+i\Gamma_\alpha]^2 -\xi_\alpha^2 -\Delta_\alpha^2}.
\label{eq.def.beta}
\end{equation}
In obtaining the last equality in Eq.~\eqref{eq.qc.DR}, we used the fact that the terms odd in $\xi_{\alpha}$ vanish upon integration.

The matrix in Eq.~\eqref{eq.qc.DR} is diagonal in the eigenbasis of $\tau_x$. We thus introduce
\begin{equation}
\ket{u_{\nu=\pm}}
=
\frac{1}{\sqrt{2}}
\begin{pmatrix}
1 \\
\pm 1
\end{pmatrix},
\end{equation}
which satisfy
\begin{equation}
\mathscr{D}^{\rm R}_{\alpha}(\omega)
\ket{u_{\nu}}
=
\left[
\beta_{\alpha}(\omega)
+
\nu\Delta_{\alpha}
\frac{\beta_{\alpha}(\omega)}
{\omega+i\Gamma_{\alpha}}
\right]
\ket{u_{\nu}}.
\end{equation}
In this eigenbasis, the $\xi_{\alpha}$-integrated Green's function
$\mathscr{D}^{\lessgtr}_{\alpha}(t-t')$ can be written as
\begin{align}
&
\mathscr{D}^{\lessgtr}_{\alpha}(t-t')=
\mp 2\sum_{\nu=\pm}
\ket{u_{\nu}}\bra{u_{\nu}}
\notag\\
&\qquad\times
\int_{-\infty}^{\infty}
\frac{d\omega}{2\pi}\,
e^{-i\omega(t-t')}
f_{\alpha}(\pm\omega)\,
\Lambda_{\alpha,\nu}(\omega),
\label{eq.qcDLG}
\end{align}
where
\begin{equation}
\Lambda_{\alpha,\nu}(\omega)
=
\operatorname{Re}\beta_{\alpha}(\omega)
+
\nu\Delta_{\alpha}
\operatorname{Re}
\left[
\frac{\beta_{\alpha}(\omega)}
{\omega+i\Gamma_{\alpha}}
\right]
\label{eq.Lambda}
\end{equation}
is the spectral factor in the $\nu$ channel.

Substituting Eq.~\eqref{eq.qcDLG} into Eq.~\eqref{eq.self.xi}, we
obtain the self-energy in the form
\begin{equation}
\bm{\Sigma}^{\lessgtr}_{\alpha}(t,t')
=
\sum_{\nu=\pm}
\m{K}^{\lessgtr}_{\alpha,\nu}(t-t')
\ket{u_{\alpha,\nu}(t)}
\bra{u_{\alpha,\nu}(t')},
\label{eq.self.Kernel}
\end{equation}
where
\begin{equation}
\ket{u_{\alpha,\nu}(t)}
=
\ket{e_{\alpha}}
\otimes
U_{\alpha}(t)
\ket{u_{\nu}},
\end{equation}
and the lesser and greater kernel functions are defined as
\begin{equation}
\m{K}^{\lessgtr}_{\alpha,\nu}(t-t')
=
i
\int_{-\infty}^{\infty}
\frac{d\omega}{2\pi}\,
e^{-i\omega(t-t')}
s^{\lessgtr}_{\alpha,\nu}(\omega).
\label{eq.kernel}
\end{equation}
Here, the functions $s^{<}_{\alpha,\nu}(\omega)$ and $s^{>}_{\alpha,\nu}(\omega)$ are given by
\begin{subequations}
\begin{align}
s^{<}_{\alpha,\nu}(\omega)
&=
2\gamma_{\alpha}
f_{\alpha}(\omega)
\Lambda_{\alpha,\nu}(\omega),
\label{eq.sl}
\\[4pt]
s^{>}_{\alpha,\nu}(\omega)
&=
-2\gamma_{\alpha}
f_{\alpha}(-\omega)
\Lambda_{\alpha,\nu}(\omega).
\label{eq.sg}
\end{align}
\end{subequations}

\subsection{Auxiliary-mode expansion}
\label{sec.AME}

The non-Markovian character of the self-energy originates from the energy dependence of the quasiparticle distribution function $f_\alpha(\omega)$ in Eq.~\eqref{eq.Fermi} and the spectral factor $\Lambda_{\alpha,\nu}(\omega)$ in Eq.~\eqref{eq.Lambda}. To handle this memory effect efficiently, we extend the auxiliary-mode expansion to the superconducting-lead self-energies derived above. This technique was originally developed for time-dependent electron transport in devices coupled to normal-metal leads~\cite{Croy2009, Popescu2016, Pavlyukh2025, Croy2011, Croy2012, Tuovinen2023, Kawamura2024, Kawamura2025}. Here, we generalize this idea to quantum systems coupled to superconducting leads.

The main idea of this technique is to evaluate the $\omega$ integral in Eq.~\eqref{eq.kernel} analytically by using the residue theorem. To this end, we approximate the distribution function $f_\alpha(\omega)$ and the spectral factor $\Lambda_{\alpha,\nu}(\omega)$ by rational functions. Specifically, we use
\begin{subequations}
\begin{align}
&
f_\alpha(\omega) 
\simeq
\frac{1}{2} +\sum_{n=1}^{N^\alpha_f}
\left[
\frac{a_{\alpha,n}}{\omega -\zeta_{\alpha,n}} +
\frac{a^*_{\alpha,n}}{\omega -\zeta^*_{\alpha,n}}
\right]
\label{eq.Fermi.AME}
,\\
&
\Lambda_{\alpha,\nu}(\omega)
\simeq
\sum_{m=1}^{N^\alpha_\Lambda}
\left[
\frac{b_{\alpha,\nu,m}}{\omega -\lambda_{\alpha,\nu,m}}
+
\frac{b^*_{\alpha,\nu,m}}{\omega -\lambda^*_{\alpha,\nu,m}}
\right].
\label{eq.lambda.AME}
\end{align}
\end{subequations}
Here, $\zeta_{\alpha,n}$ and $\lambda_{\alpha,\nu,m}$ are poles in the upper half of the complex plane, satisfying ${\rm Im}\,\zeta_{\alpha,n}>0$ and ${\rm Im}\,\lambda_{\alpha,\nu,m}>0$, and $a_{\alpha,n}$ and $b_{\alpha,\nu,m}$ are the corresponding residues. The conjugate poles in the lower half of the complex plane are included so that the approximated functions are real on the real-frequency axis.

A key practical issue in using Eqs.~\eqref{eq.Fermi.AME} and \eqref{eq.lambda.AME} is how to choose the poles and residues so as to approximate the original functions accurately with as few modes as possible. For the distribution function $f_\alpha(\omega)$, we use the Pad\'e decomposition~\cite{Ozaki2007, Hu2010, Karrasch2010}, which provides a rapidly convergent and numerically stable rational approximation. In this method, the problem of determining the poles $\zeta_{\alpha,n}$ and residues $a_{\alpha,n}$ in Eq.~\eqref{eq.Fermi.AME} is mapped onto an eigenvalue problem of a real symmetric tridiagonal matrix. The matrix $S$ used for this purpose is defined by its nonzero elements,
\begin{equation}
S_{j,\, j+1}=
S_{j+1,\, j}=
\frac{1}{2\sqrt{(2j-1)(2j+1)}}.
\end{equation}
Here $1\le j\le 2N^\alpha_f-1$. Because of the symmetry of $S$, its eigenvalues appear in pairs $(s_n,-s_n)$. We take the positive eigenvalues $s_n>0$ $(n=1,\cdots,N^\alpha_f)$ and denote the corresponding normalized eigenvectors by $\ket{s_n}$. The poles and residues are then given by
\begin{equation}
\zeta_{\alpha,n}=
i\frac{T_{{\rm SC},\alpha}}{s_n},
\qquad
a_{\alpha,n}
=
-T_{{\rm SC},\alpha}
\frac{|\braket{1|s_n}|^2}{4s_n^2},
\end{equation}
where $\bra{1}=(1,0,\cdots,0)$. In this scheme, the residues $a_{\alpha,n}$ are real. The number $N^\alpha_f$ of poles required for an accurate approximation depends on the temperature $T_{{\rm SC},\alpha}$ and the relevant energy window. In typical calculations, however, we find that $N^\alpha_f\simeq 10-30$ is sufficient to achieve good accuracy.

For the spectral factor $\Lambda_{\alpha,\nu}(\omega)$, we determine the poles $\lambda_{\alpha,\nu,m}$ and residues $b_{\alpha,\nu,m}$ in Eq.~\eqref{eq.lambda.AME} using the adaptive Antoulas–Anderson (AAA) algorithm~\cite{Nakatsukasa2018}. The AAA algorithm constructs a rational approximation to a given function from its values sampled on the real-frequency axis. Starting from a set of sampling points, it adaptively selects support points so as to reduce the approximation error. The resulting rational approximant is then converted into the pole-residue form in Eq.~\eqref{eq.lambda.AME}. This procedure is well suited to the present problem because $\Lambda_{\alpha,\nu}(\omega)$ has a nontrivial energy dependence originating from the superconducting density of states and the anomalous coherence factors.

\begin{figure}[t]
\centering
\includegraphics[width=\linewidth]{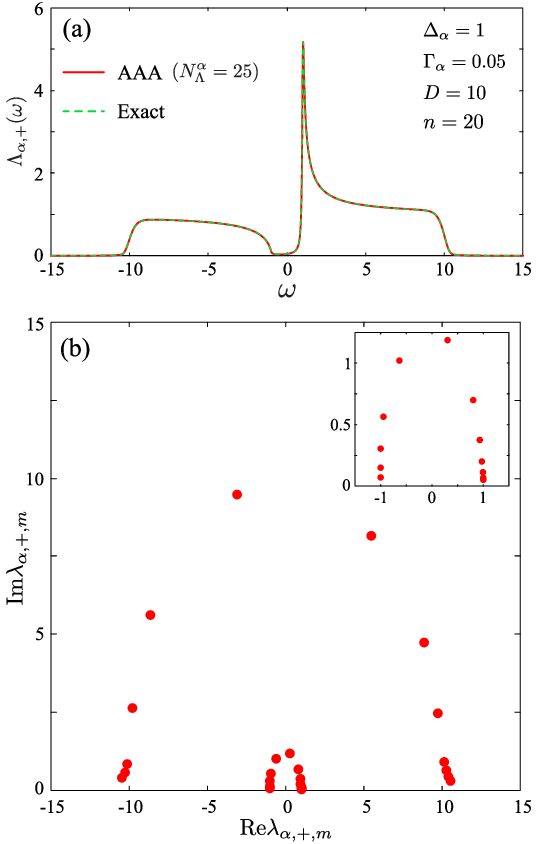}
\caption{
Rational approximation of the spectral factor $\Lambda_{\alpha,+}(\omega)$ obtained by the AAA algorithm. The parameters are $\Delta_\alpha=1$, $D=10$, $\Gamma_\alpha=0.05$, and $n=20$ in the cutoff function. (a) Comparison between the exact result for $\Lambda_{\alpha,+}(\omega)$ and its rational approximation in Eq.~\eqref{eq.lambda.AME} with $N^\alpha_\Lambda=25$ poles. The solid and dashed lines represent the approximate and exact results, respectively. The approximation error satisfies $|\Lambda^{\rm exact}_{\alpha,+}(\omega)-\Lambda^{\rm AAA}_{\alpha,+}(\omega)|<10^{-5}$ for all $\omega\in[-15,15]$. (b) Positions of the AAA poles $\lambda_{\alpha,+,m}$ used in the rational approximation shown in panel (a). The poles are plotted in the upper half of the complex plane. The inset shows a magnified view near the origin.}
\label{Fig_AAA} 
\end{figure}

Figure~\ref{Fig_AAA} shows a typical example of the rational approximation of $\Lambda_{\alpha,\nu}(\omega)$ obtained by the AAA algorithm. In Fig.~\ref{Fig_AAA}(a), we compare the exact function $\Lambda_{\alpha,+}(\omega)$ with its rational approximation in Eq.~\eqref{eq.lambda.AME}, using $N^\alpha_\Lambda=25$ poles. Figure~\ref{Fig_AAA}(b) shows the positions of the corresponding poles $\lambda_{\alpha,+,m}$ in the upper half of the complex plane. We find that the rational approximation reproduces the original spectral factor with high accuracy using only a small number of poles.

Once the poles and residues have been determined, the frequency integral in Eq.~\eqref{eq.kernel} can be evaluated analytically by the residue theorem. Substituting the rational approximations in Eqs.~\eqref{eq.Fermi.AME} and \eqref{eq.lambda.AME} into Eq.~\eqref{eq.kernel}, we obtain the mode expansion
\begin{subequations}
\begin{align}
&
\m{K}^\lessgtr_{\alpha,\nu}(t -t')=
\sum_{l=1}^{N^\alpha_{\rm pole}}
\m{K}^\lessgtr_{\alpha,\nu,l}(t -t')
\label{eq.Kernel.AME}
,\\
&
\m{K}^\lessgtr_{\alpha,\nu,l}(t -t')=
\begin{cases}
\big(R^\lessgtr_{\alpha,\nu,l}\big)^*\,
e^{-i\chi^*_{\alpha,\nu,l}(t-t')}
\hspace{0.3cm} (t >t')
\\[6pt]
-R^\lessgtr_{\alpha,\nu,l}\,
e^{-i\chi_{\alpha,\nu,l}(t - t')}
\hspace{0.55cm} (t <t')
\end{cases}.
\label{eq.Kernel.AME.def}
\end{align}
\end{subequations}
Here, we have introduced a single pole index $l$ by combining the poles originating from the Pad\'e expansion of $f_\alpha(\omega)$ and those obtained from the AAA approximation of $\Lambda_{\alpha,\nu}(\omega)$. Thus,
\begin{equation}
N^\alpha_{\rm pole}=
N^\alpha_f+N^\alpha_\Lambda .
\end{equation}
With this convention, the exponents $\chi_{\alpha,\nu,l}$ and the coefficients $R^{\lessgtr}_{\alpha,\nu,l}$ are defined by
\begin{subequations}
\begin{align}
&
\chi_{\alpha,\nu,l}=
\Big\{
\zeta_{\alpha,n},\quad \lambda_{\alpha,\nu,m}
\Big\}
\label{eq.def.chi}
,\\[4pt]
&
R^<_{\alpha,\nu,l} =
\Big\{
2\gamma_\alpha a_{\alpha,n} \Lambda(\zeta_{\alpha,n}),\quad
2\gamma_\alpha b_{\alpha,\nu,m} f(\lambda_{\alpha,\nu,m})
\Big\}
\label{eq.def.Rl}
,\\
&
R^>_{\alpha,\nu,l} =
\Big\{
2\gamma_\alpha a_{\alpha,n} \Lambda(\zeta_{\alpha,n}),\quad
-2\gamma_\alpha b_{\alpha,\nu,m} f(-\lambda_{\alpha,\nu,m})
\Big\}. 
\label{eq.def.Rg}
\end{align}
\end{subequations}
Using the expanded kernel in Eq.~\eqref{eq.Kernel.AME}, the self-energy in Eq.~\eqref{eq.self.Kernel} is decomposed into auxiliary modes as
\begin{subequations}
\begin{align}
&
\bm{\Sigma}^\lessgtr_\alpha(t, t')=
\sum_{\nu=\pm} \sum_{l=1}^{N^\alpha_{\rm pole}}
\bm{\Sigma}^\lessgtr_{\alpha, \nu, l}(t, t')
\label{eq.self.AME}
,\\[4pt]
&
\bm{\Sigma}^\lessgtr_{\alpha, \nu, l}(t, t')
=
\m{K}^\lessgtr_{\alpha,\nu,l}(t -t')
\ket{u_{\alpha,\nu}(t)}
\bra{u_{\alpha,\nu}(t')}.
\label{eq.self.AME.2}
\end{align}
\end{subequations}
This form is the starting point for converting the memory integral in the KB equation into time-local EOMs for auxiliary variables.

Before proceeding to the derivation of the time-local EOMs, we comment on the roles of the Dynes parameter $\Gamma_\alpha$ and the cutoff function $W_{\rm c}(\xi_\alpha)$ introduced in Eqs.~\eqref{eq.DR} and~\eqref{eq.ksum.cut}, respectively.

A nonzero $\Gamma_\alpha$ is essential for the convergence of the auxiliary-mode expansion. In the limit $\Gamma_\alpha \to 0^+$, the square-root singularities of $\Lambda_{\alpha,\nu}(\omega)$ at $\omega=\pm\Delta_\alpha$ lie on the real axis, giving rise to a power-law memory kernel that cannot be represented by a finite sum of exponential modes. A nonzero $\Gamma_\alpha$ shifts these singularities away from the real axis, ensuring convergence of the rational approximation in Eq.~\eqref{eq.lambda.AME}.

The cutoff function $W_{\rm c}(\xi_\alpha)$ is introduced to
keep the formulation transparent. If one takes the wide-band limit, $W_{\rm c}(\xi_\alpha)=1$, the rational approximation of $\Lambda_{\alpha,\nu}(\omega)$ in Eq.~\eqref{eq.lambda.AME} generally contains a constant contribution associated with the high-energy limit $|\omega|\to\infty$. Together with the constant term $1/2$ in Eq.~\eqref{eq.Fermi.AME}, this contribution generates an additional term proportional to $\delta(t-t')$ in the kernel $\m{K}^\lessgtr_{\alpha,\nu}(t-t')$. Although such a term can be treated in principle, it makes the subsequent derivation considerably more cumbersome and leads to a less transparent set of time-local EOMs. Since the transport dynamics of interest is governed by the energy window around the Fermi level, discarding these high-energy contributions
does not affect the low-energy dynamics.

\subsection{Derivation of Eqs.~\eqref{eq.EOM.Pi} and \eqref{eq.EOM.Omega}}

We now derive Eqs.~\eqref{eq.EOM.Pi} and \eqref{eq.EOM.Omega}. We start from the current matrix $\bm{\Pi}_\alpha(t)$ defined in Eq.~\eqref{eq.Pi}. Substituting the mode-expanded self-energy in Eq.~\eqref{eq.self.AME}, we decompose $\bm{\Pi}_\alpha(t)$ into auxiliary modes as
\begin{align}
\bm{\Pi}_\alpha(t) 
&=
\sum_{\nu=\pm} 
\sum_{l=1}^{N^\alpha_{\rm pole}}
\int_{-\infty}^t dt_1\,
\big[
\bm{G}^>(t,t_1) \bm{\Sigma}^<_{\alpha,\nu, l}(t_1,t)
\notag\\
&\hspace{2cm}-
\bm{G}^<(t,t_1) \bm{\Sigma}^>_{\alpha,\nu, l}(t_1,t)
\big]
\notag\\
& \equiv
\sum_{\nu=\pm}
\sum_{l=1}^{N^\alpha_{\rm pole}}\,
\bm{\Pi}_{\alpha, \nu, l}(t).
\label{eq.Pi.AME}
\end{align}
Taking the time derivative of $\bm{\Pi}_{\alpha,\nu,l}(t)$, one obtains
\begin{align}
i\frac{d}{dt}\, \bm{\Pi}_{\alpha,\nu,l}(t)
&=
-R^<_{\alpha,\nu,l}\, \bm{P}_{\alpha,\nu}(t) 
\notag\\[4pt]
&\qquad
+
i\big[R^>_{\alpha,\nu,l} - R^<_{\alpha,\nu,l}\big]
\bm{G}^<(t)\, \bm{P}_{\alpha,\nu}(t) 
\notag\\[4pt]
&\qquad
+
\big[\bm{h}_{\rm sys}(t)- \chi_{\alpha,\nu,l}\big]
\bm{\Pi}_{\alpha,\nu,l}(t)
\notag\\[4pt]
&\qquad
-
\bm{\Pi}_{\alpha,\nu,l}(t)\,
\bm{\Phi}_\alpha(t)
+
\bm{\Omega}_{\alpha,\nu,l}(t).
\label{eq.dt.Pi}
\end{align}
This equation is equivalent to Eq.~\eqref{eq.EOM.Pi} after flattening the indices $(\alpha,\nu,l)$ into the composite index $K$. In Eq.~\eqref{eq.dt.Pi}, the time-dependent matrix $\bm{P}_{\alpha,\nu}(t)$ is defined by
\begin{equation}
\bm{P}_{\alpha,\nu}(t) =
\ket{u_{\alpha,\nu}(t)}
\bra{u_{\alpha,\nu}(t)},
\label{eq.def.P}
\end{equation}
and the superconducting phase-velocity matrix is given by
\begin{equation}
\bm{\Phi}_\alpha(t)=
I_N \otimes
\frac{1}{2}\, \frac{d\phi_\alpha(t)}{dt}\, \tau_z.
\label{eq.def.Phi.mat}
\end{equation}
The first two terms on the right-hand side of Eq.~\eqref{eq.dt.Pi} originate from the upper limit of the time integral in Eq.~\eqref{eq.Pi.AME}. The third and fourth terms describe the time evolution generated by the central-system Hamiltonian and by the time-dependent superconducting phase, respectively. The last term, $\bm{\Omega}_{\alpha,\nu,l}(t)$, contains the remaining memory contribution. The detailed derivation of Eq.~\eqref{eq.dt.Pi} is given in Appendix~\ref{sec.app.dt.Pi}.

The auxiliary matrix $\bm{\Omega}_{\alpha,\nu,l}(t)$ is defined as
\begin{align}
\bm{\Omega}_{\alpha,\nu,l}(t)
&=
\int_{-\infty}^\infty dt_1\,
\int_{-\infty}^\infty dt_2\,
\Big[
\bm{\Sigma}^{\rm R}(t, t_2)
\bm{G}^>(t_2,t_1)
\notag\\
&\hspace{1cm}
+
\bm{\Sigma}^>(t, t_2)
\bm{G}^{\rm A}(t_2,t_1)
\Big]
\bm{\Sigma}^<_{\alpha,\nu, l}(t_1,t)
\notag\\[4pt]
&\quad
-
\int_{-\infty}^\infty dt_1\,
\int_{-\infty}^\infty dt_2\,
\Big[
\bm{\Sigma}^{\rm R}(t, t_2)
\bm{G}^<(t_2,t_1)
\notag\\
&\hspace{1cm}
+
\bm{\Sigma}^<(t, t_2)
\bm{G}^{\rm A}(t_2,t_1)
\Big]
\bm{\Sigma}^>_{\alpha,\nu, l}(t_1,t).
\label{eq.def.Omega}
\end{align}
Here, the total lesser and greater self-energies are
\begin{equation}
\bm{\Sigma}^{\lessgtr}(t,t')
=
\sum_{\alpha=1}^{N_{\rm lead}}
\bm{\Sigma}^{\lessgtr}_\alpha(t,t'),
\end{equation}
and the retarded self-energy and advanced Green's function are related to the lesser and greater components by~\cite{Datta1995, Haug2008, Stefanucci2013}
\begin{subequations}
\begin{align}
&
\bm{\Sigma}^{\rm R}(t,t') = 
\Theta(t-t') \big[\bm{\Sigma}^>(t,t') -\bm{\Sigma}^<(t,t') \big]
\label{eq.self.R}
,\\[2pt]
&
\bm{G}^{\rm A}(t,t') = 
-\Theta(t' -t) \big[\bm{G}^>(t,t') -\bm{G}^<(t,t') \big],
\label{eq.GA}
\end{align}
\end{subequations}
where $\Theta(t)$ is the step function.

Equation~\eqref{eq.dt.Pi} is not yet closed, because it contains the auxiliary matrix $\bm{\Omega}_{\alpha,\nu,l}(t)$. By substituting the mode expansion of the self-energy into Eq.~\eqref{eq.def.Omega}, this quantity can be further decomposed into contributions from pairs of auxiliary modes. Introducing the composite indices
\begin{equation}
K= \{ \alpha, \nu, l\}
,\quad
K'= \{ \alpha', \nu', l'\},
\end{equation}
we write
\begin{align}
\bm{\Omega}_{\alpha,\nu,l}(t)
&=
\sum_{\alpha'=1}^{N_{\rm lead}}\sum_{\nu'=\pm}\sum_{l'=1}^{N^{\alpha'}_{\rm pole}}\, \bm{\Omega}_{\alpha,\nu,l,\alpha',\nu',l'}(t)
\notag\\
&=
\sum_{K'} \bm{\Omega}_{K, K'}(t).
\label{eq.Omega.AME}
\end{align}
Here, $\bm{\Omega}_{K,K'}(t)$ is defined by
\begin{widetext}
\begin{align}
\bm{\Omega}_{K, K'}(t)
&=
\int_{-\infty}^t dt_1
\int_{-\infty}^t dt_2\,
\Big[
\bm{\Sigma}^>_{K'}(t, t_2) -
\bm{\Sigma}^<_{K'}(t, t_2)
\Big]
\Big[
\bm{G}^>(t_2,t_1) \bm{\Sigma}^<_K(t_1,t) -
\bm{G}^<(t_2,t_1) \bm{\Sigma}^>_K(t_1,t)
\Big]
\notag\\
&\qquad
-
\int_{-\infty}^t dt_1
\int_{-\infty}^{t_1} dt_2\,
\Big[
\bm{\Sigma}^<_{K'}(t, t_2)
\bm{G}^{\rm A}(t_2,t_1)
\bm{\Sigma}^>_{K}(t_1,t)
-
\bm{\Sigma}^>_{K'}(t, t_2)
\bm{G}^{\rm A}(t_2,t_1)
\bm{\Sigma}^<_{K}(t_1,t)
\Big].
\label{eq.Omega.KK}
\end{align}
\end{widetext}
Taking the time derivative of $\bm{\Omega}_{K,K'}(t)$ yields the EOM for $\bm{\Omega}_{K,K'}(t)$ given in Eq.~\eqref{eq.EOM.Omega}. The calculation is straightforward but lengthy and is thus presented in Appendix~\ref{sec.app.EOM.Omega}.

\subsection{Initial condition at $t=0$}
\label{sec.Initial}

Lastly, we specify the initial conditions for the coupled time-local EOMs in Eqs.~\eqref{eq.EOM.GL}–\eqref{eq.EOM.Omega}. As explained in Sec.~\ref{sec.Hamiltonian}, all time-dependent perturbations are switched off for $t<0$, and the coupled system is assumed to be in thermal equilibrium with time-independent superconducting phases $\phi_\alpha(0)$. The initial equal-time lesser Green's function is thus obtained from the equilibrium lesser Green's function as
\begin{equation}
\bm{G}^<(t=0) = 
\int_{-\infty}^\infty \frac{d\omega}{2\pi}\,
\bm{G}^<_{\rm eq}(\omega).
\label{eq.GL.ini}
\end{equation}
The equilibrium lesser Green's function is calculated from the Keldysh equation~\cite{Datta1995, Haug2008, Stefanucci2013}
\begin{equation}
\bm{G}^<_{\rm eq}(\omega)=
\bm{G}^{\rm R}_{\rm eq}(\omega)
\bm{\Sigma}^<_{\rm eq}(\omega)
\bm{G}^{\rm A}_{\rm eq}(\omega),
\label{eq.GL.ini.Dyson}
\end{equation}
where the retarded and advanced Green's functions are given by~\cite{Datta1995, Haug2008, Stefanucci2013}
\begin{equation}
\bm{G}^{\rm R(A)}_{\rm eq}(\omega) =
\frac{1}{\omega -\bm{h}_{\rm sys}(0) -\bm{\Sigma}^{\rm R(A)}_{\rm eq}(\omega)}.
\label{eq.GR.ini}
\end{equation}
Here, $\bm{\Sigma}^{\rm R(A)}_{\rm eq}(\omega)$ and $\bm{\Sigma}^<_{\rm eq}(\omega)$ are the equilibrium self-energies due to the superconducting leads, evaluated with the time-independent phases $\phi_\alpha(0)$.

In Eqs.~\eqref{eq.GL.ini.Dyson} and \eqref{eq.GR.ini}, the equilibrium self-energies are given by
\begin{subequations}
\begin{align}
&
\bm{\Sigma}^{\rm R}_{{\rm eq},\alpha}(\omega) 
=
-i\gamma_\alpha\,
\beta_\alpha(\omega)
\ket{e_\alpha}\bra{e_\alpha}
\otimes
\scalebox{0.95}{$\displaystyle
\begin{pmatrix}
1& 
\dfrac{\Delta_\alpha(0)}{\omega +i\Gamma_\alpha} \\[4pt]
\dfrac{\Delta^*_\alpha(0)}{\omega +i\Gamma_\alpha} &
1
\end{pmatrix}
$},\\
&
\bm{\Sigma}^<_{{\rm eq},\alpha}(\omega)
=
2i\gamma_\alpha f_\alpha(\omega)
\ket{e_\alpha}\bra{e_\alpha}
\notag\\
&\qquad
\otimes
\begin{pmatrix}
{\rm Re}\, \beta_\alpha(\omega) 
&
\Delta_\alpha(0)\, 
{\rm Re}\left[\dfrac{\beta_\alpha(\omega)}{\omega +i\Gamma_\alpha}\right]
\\[8pt]
\Delta^*_\alpha(0)\, 
{\rm Re}\left[\dfrac{\beta_\alpha(\omega)}{\omega +i\Gamma_\alpha}\right]
&
{\rm Re}\, \beta_\alpha(\omega)
\end{pmatrix}.
\end{align}
\end{subequations}
Here, we have explicitly carried out the multiplication by the phase matrices $U_\alpha(0)$ and $U_\alpha^\dagger(0)$. The resulting phase dependence is included in the initial pair potential,
\begin{equation}
\Delta_\alpha(0) \equiv \Delta_\alpha\, e^{i\phi_\alpha(0)}.
\end{equation}
The total equilibrium self-energies are obtained by summing over the leads,
\begin{equation}
\bm{\Sigma}^{\rm R,<}_{\rm eq}(\omega)
=
\sum_{\alpha=1}^{N_{\rm lead}}
\bm{\Sigma}^{\rm R,<}_{{\rm eq},\alpha}(\omega),
\end{equation}
and $\bm{\Sigma}^{\rm A}_{\rm eq}(\omega)=\big[\bm{\Sigma}^{\rm R}_{\rm eq}(\omega)\big]^\dagger$. Using these self-energies in Eqs.~\eqref{eq.GL.ini.Dyson} and \eqref{eq.GR.ini}, we numerically evaluate the initial lesser Green's function $\bm{G}^<(0)$ in Eq.~\eqref{eq.GL.ini}.

We next determine the initial condition for the mode-resolved current matrix $\bm{\Pi}_K(t)$. It is useful to rewrite Eq.~\eqref{eq.Pi.AME} in terms of the retarded Green's function and the advanced self-energy as
\begin{align}
\bm{\Pi}_K(t)
&=
\int_{-\infty}^{\infty} dt_1\,
\big[
\bm{G}^{\rm R}(t,t_1) \bm{\Sigma}^<_K(t_1,t)
\notag\\
&\hspace{3cm}
+
\bm{G}^<(t,t_1) \bm{\Sigma}^{\rm A}_K(t_1,t)
\big].
\label{eq.PiK.AME.2}	
\end{align}
Although the original expression in Eq.~\eqref{eq.Pi.AME} contains an integral over $-\infty<t_1<t$, the integration range can be extended to the whole time axis in Eq.~\eqref{eq.PiK.AME.2}, because the retarded and advanced components contain the corresponding step functions. This form is convenient for evaluating the initial condition. Fourier transforming Eq.~\eqref{eq.PiK.AME.2}, we obtain
\begin{equation}
\bm{\Pi}_{K}(0)
=
\int_{-\infty}^{\infty} \frac{d\omega}{2\pi}\,
\big[
\bm{G}^{\rm R}_{\rm eq}(\omega) 
\bm{\Sigma}^<_{{\rm eq}, K}(\omega)
+
\bm{G}^<_{\rm eq}(\omega) 
\bm{\Sigma}^{\rm A}_{{\rm eq}, K}(\omega)
\big].
\label{eq.PiK.ini}
\end{equation}
The mode-resolved equilibrium self-energies are obtained from Eqs.~\eqref{eq.Kernel.AME.def} and \eqref{eq.self.AME.2} as
\begin{subequations}
\begin{align}
\bm{\Sigma}^<_{{\rm eq}, K}(\omega)
&=
2i\, {\rm Re}\left[\frac{R^<_{K}}{\omega -\chi_{K}}\right]
\bm{P}_{K}(0)
\label{eq.SigL.ini}
,\\[4pt]
\bm{\Sigma}^{\rm A}_{{\rm eq}, K}(\omega)
&=
-i\, 
\frac{R^>_K - R^<_K}{\omega -\chi_K}\bm{P}_{K}(0).
\label{eq.SigA.ini}
\end{align}
\end{subequations}
Here, the matrix $\bm{P}_{K}(0)=\bm{P}_{\alpha,\nu}(0)$ is independent of the pole index $l$ and is explicitly given by
\begin{equation}
\bm{P}_{\alpha,\nu=\pm}(0)
=
\ket{e_\alpha}\bra{e_\alpha}
\otimes
\frac{1}{2}
\begin{pmatrix}
1 & \nu e^{-i\phi_\alpha(0)} \\[4pt]
\nu e^{i\phi_\alpha(0)} & 1
\end{pmatrix}.
\label{eq.P.ini}
\end{equation}
Substituting Eqs.~\eqref{eq.GL.ini.Dyson}, \eqref{eq.GR.ini}, \eqref{eq.SigL.ini}, and \eqref{eq.SigA.ini} into Eq.~\eqref{eq.PiK.ini}, we numerically evaluate the initial current matrix $\bm{\Pi}_{K}(0)$.

To complete the initial conditions, we determine the initial auxiliary matrix $\bm{\Omega}_{K,K'}(0)$. Although it can be evaluated directly from Eq.~\eqref{eq.Omega.KK}, it is more convenient to construct it from the initial current matrix $\bm{\Pi}_K(0)$. In the initial equilibrium state, all superconducting phases are time independent, and hence $\bm{\Phi}_{K}(0)=0$. Moreover, the auxiliary variables are stationary for $t<0$. Setting $\partial_t\, \bm{\Omega}_{K,K'}(t)=0$ in Eq.~\eqref{eq.EOM.Omega}, we obtain
\begin{align}
\bm{\Omega}_{K,K'}(0)
&=
\frac{i}{\chi^*_{K'} - \chi_{K}}
\bigg[
\big[R^>_{K} -R^<_{K}\big]\, 
\bm{\Pi}^\dagger_{K'}(0)\,
\bm{P}_{K}(0)
\notag\\
&\qquad
-
\big[\big(R^>_{K'}\big)^* -\big(R^<_{K'}\big)^*\big]\,
\bm{P}_{K'}(0)\,
\bm{\Pi}_{K}(0)
\bigg].
\label{eq.Omega.ini}	
\end{align}
We thus solve the coupled time-local EOMs~\eqref{eq.EOM.GL}–\eqref{eq.EOM.Omega} with the initial conditions given in Eqs.~\eqref{eq.GL.ini}, \eqref{eq.PiK.ini}, and \eqref{eq.Omega.ini}.

\section{Application: voltage-quench dynamics in an S-QD-S junction}
\label{sec.application}

In this section, we apply the time-local formulation developed in Sec.\ref{sec.Formalism} to a minimal superconductor–quantum-dot–superconductor (S-QD-S) junction. This system provides a useful benchmark for the present real-time scheme, because the long-time dynamics after a voltage quench is expected to approach the ac Josephson periodic steady state. This periodic steady state can be independently obtained from the Floquet Green's function method~\cite{Sun2002}, allowing us to test the validity of the proposed time-local EOMs.

We emphasize that no periodic ansatz is imposed in the present real-time calculation. Starting from the equilibrium state, the system evolves under a suddenly applied dc bias and eventually reaches a periodic steady state with the Josephson frequency. The transient regime before this relaxation is also directly accessible within our framework. In particular, we show an example in which the early-time current exhibits a subharmonic component, consistent with the nonequilibrium fractional Josephson effect proposed in recent work~\cite{Lahiri2023}.

\subsection{Model of the S-QD-S junction}
\label{sec.app.model}

We consider a single-level, spin-degenerate quantum dot (QD) coupled to two conventional $s$-wave superconducting leads, labeled by $\alpha=1,2$. Electron–electron interactions on the QD are neglected in this benchmark calculation. This setup corresponds to the $N=1$ case of the general formulation in Sec.\ref{sec.Hamiltonian}. Thus, the $2N$-component Nambu field reduces to the two-component QD Nambu spinor
\begin{equation}
\Psi = 
\Psi_1 =
\begin{pmatrix}
c_{\up} \\[4pt]
c^\dagger_\down
\end{pmatrix},
\end{equation}
where $c_\sigma$ annihilates an electron with spin $\sigma$ on the dot. The system Hamiltonian in Eq.~\eqref{eq.Hsys} is then specified as
\begin{equation}
H_{\rm sys}(t)
=
\Psi^\dagger\, \bm{h}_{\rm sys}(t)\, \Psi,
\end{equation}
with the $2\times 2$ single-particle Hamiltonian
\begin{equation}
\bm{h}_{\rm sys} = \ep_{\rm d}(t)\, \tau_z.
\end{equation}
Here $\epsilon_{\rm d}(t)$ denotes the QD energy level. In the numerical results below, $\epsilon_{\rm d}(t)$ is taken to be a time-independent constant, although the formalism itself allows general time-dependent one-body potentials in the QD.

In the present S-QD-S setup, the two superconducting leads are attached to the same QD orbital. Since $N=1$, the channel vector in Eq.~\eqref{eq.V} reduces to
\begin{equation}
\ket{e_{\alpha}} = 1
\quad
(\alpha=1,2).
\end{equation}
We also take the two superconducting leads to have the same gap amplitude, temperature, and Dynes broadening parameter,
\begin{subequations}
\begin{align}
& \Delta_1=\Delta_2\equiv \Delta,\\[2pt]
& T_{{\rm SC},1}=T_{{\rm SC},2}\equiv T_{\rm SC},\\[2pt]
& \Gamma_1=\Gamma_2\equiv \Gamma.
\end{align}
\end{subequations}
The coupling strength between the QD and lead $\alpha$ is characterized by $\gamma_\alpha$, defined in Eq.~\eqref{eq.gamma.def}.  In the following, we focus on symmetric junctions with $\gamma_1=\gamma_2$ ($\equiv \gamma$) for simplicity.

To study the relaxation toward the ac Josephson periodic steady state, we consider a voltage quench. The system is initially prepared in an equilibrium state for $t\leq 0$, where the two superconducting phases are fixed at $\phi_1(0)=\phi_2(0)=0$. At $t=0$, a dc voltage bias is switched on as
\begin{equation}
V_1(t)=\frac{V}{2}\Theta(t),
\quad
V_2(t)=-\frac{V}{2}\Theta(t).
\label{eq.app.voltage.quench}
\end{equation}
In this voltage-quench protocol, Eq.~\eqref{eq.phi.V} gives, for $t>0$,
\begin{equation}
\phi_1(t)=eVt,
\quad
\phi_2(t)=-eVt.
\label{eq.phi.application}
\end{equation}
Thus the phase difference evolves as
\begin{equation}
\phi_1(t)-\phi_2(t)
= \omega_J t,
\end{equation}
where $\omega_J=2eV$ is the Josephson frequency~\cite{Josephson1962, TinkhamBook}.

In the present $N=1$ case, both $\bm{P}_{\alpha,\nu}(t)$ in Eq.~\eqref{eq.def.P} and $\bm{\Phi}_{\alpha}(t)$ in Eq.~\eqref{eq.def.Phi.mat} reduce to $2\times2$ matrices in Nambu space. They are explicitly given by
\begin{subequations}
\begin{align}
\bm{P}_{\alpha,\nu}(t)
&=
\frac{1}{2}
\begin{pmatrix}
1 & \nu\, e^{-i\phi_\alpha(t)} \\
\nu\, e^{i\phi_\alpha(t)} & 1
\end{pmatrix},
\label{eq.app.P.matrix}
\\[4pt]
\bm{\Phi}_{\alpha}(t)
&=
\frac{1}{2}\frac{d\phi_\alpha(t)}{dt}\tau_z .
\label{eq.app.Phi.matrix}
\end{align}
\end{subequations}
For the voltage-quench protocol in Eq.~\eqref{eq.app.voltage.quench}, one has $\dot{\phi}_{1,2}(t) = \pm eV$ for $t>0$.

Although we focus on the step-like voltage quench in Eq.~\eqref{eq.app.voltage.quench} as a minimal protocol, the time-local formulation itself does not rely on this specific choice. More general voltage protocols, such as pulsed, ramped, or periodically modulated biases, can be treated by substituting the corresponding $V_\alpha(t)$ into Eq.~\eqref{eq.phi.V}.

With these specifications, the dynamics of the S-QD-S junction is obtained by solving the time-local EOMs~\eqref{eq.EOM.GL}–\eqref{eq.EOM.Omega}. Since the present setup is the $N=1$ specialization of the general formulation, we do not repeat the full equations here. The dynamical variables to be propagated are
\begin{equation}
\bm{G}^{<}(t),
\quad
\bm{\Pi}_{K}(t),
\quad
\bm{\Omega}_{K,K'}(t),
\end{equation}
where $K=\{\alpha,\nu,l\}$ denotes the composite index for the lead, the Nambu branch, and the auxiliary pole. In the present single-level QD case, all these quantities are $2\times2$ matrices in Nambu space. The model-dependent inputs to the EOMs are $\bm{h}_{\rm sys}(t)$, $\bm{P}_{\alpha,\nu}(t)$, and $\bm{\Phi}_{\alpha}(t)$, which have been specified above. The time-dependent current from lead $\alpha$ is evaluated using Eq.~\eqref{eq.current.J}. In the present $N=1$ case, it is given by
\begin{equation}
J_\alpha(t)
=
2e\,{\rm Re}\,
{\rm Tr}
\left[
\tau_z
\sum_{\nu=\pm} \sum_{l=1}^{N^\alpha_{\rm pole}}
\bm{\Pi}_{\alpha,\nu,l}(t)
\right].
\label{eq.app.current}
\end{equation}

\subsection{Transient dynamics and relaxation to the ac Josephson steady state}
\label{sec.app.results}

\begin{figure}[t]
\centering
\includegraphics[width=\linewidth]{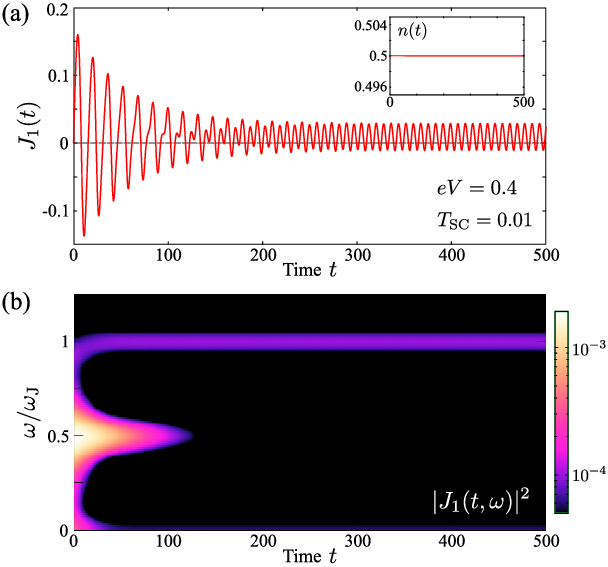}
\caption{Voltage-quench dynamics in the subgap-bias regime for $eV=0.4$ and $T_{\rm SC}=0.01$. (a) Time evolution of the current $J_1(t)$ flowing from superconducting lead 1 into the QD after the voltage quench. The inset shows the corresponding time evolution of the dot occupation $n(t)$. (b) Local spectrum $|J_1(t,\omega)|^2$ obtained by applying a short-time Fourier transform to $J_1(t)$ in panel (a). The frequency $\omega$ is plotted in units of the Josephson frequency $\omega_{\rm J}=2eV$.
}
\label{Fig_ACJE} 
\end{figure}

In the following numerical calculations, we use the superconducting gap $\Delta$ as the unit of energy and set $\Delta=1$. Unless otherwise stated, the coupling strengths are chosen as $\gamma=0.1$, the Dynes broadening parameter in the superconducting leads is set to $\Gamma=0.05$, and the bandwidth is taken to be $D=10$. The dot level is fixed at the time-independent particle–hole symmetric point, $\epsilon_{\rm d}(t)=0$, throughout this section. In the auxiliary-mode expansion, we use $N_f=15$ poles for the Fermi distribution in Eq.~\eqref{eq.Fermi.AME} and $N_\Lambda=25$ poles for the spectral factor in Eq.~\eqref{eq.lambda.AME}. We have confirmed that the results presented below are well converged with respect to the number of poles.

We first examine the voltage-quench dynamics in a typical subgap-bias regime, $eV<2\Delta$. Figure~\ref{Fig_ACJE} shows the results for $T_{\rm SC}=0.01$ and $eV=0.4$. Figure~\ref{Fig_ACJE}(a) shows the time evolution of the current $J_1(t)$ flowing from superconducting lead 1 into the QD. The inset shows the dot occupation, $n_\uparrow(t)=n_\downarrow(t)\equiv n(t)$. In the symmetric setup considered here, particle–hole symmetry keeps $n(t)=0.5$ throughout the time evolution, and the relation $J_1(t)=-J_2(t)$ is always satisfied. To resolve the time-dependent frequency components of the transient current, Fig.~\ref{Fig_ACJE}(b) shows the local spectrum $|J_1(t,\omega)|^2$, obtained by applying a short-time Fourier transform to $J_1(t)$ in Fig.~\ref{Fig_ACJE}(a). In this paper, we define the short-time Fourier transform as
\begin{equation}
J_1(t,\omega)
=
\int dt'\,
e^{i\omega(t'-t)}
\exp\left[-\frac{(t'-t)^2}{2\sigma^2}\right]
J_1(t'),
\end{equation}
where we use a Gaussian window with width $\sigma=20$.

\begin{figure}[t]
\centering
\includegraphics[width=\linewidth]{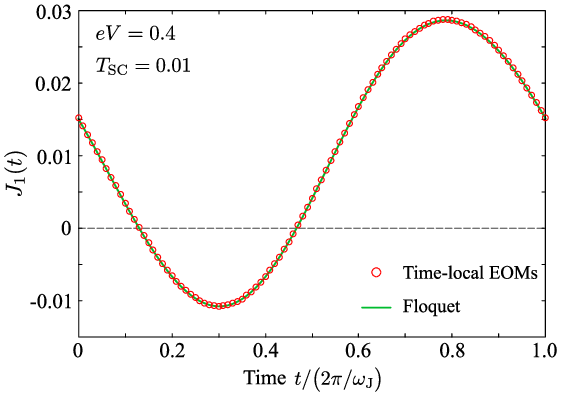}
\caption{
Comparison of the ac Josephson periodic steady state for $eV=0.4$ and $T_{\rm SC}=0.01$. The figure shows one period of the current $J_1(t)$ in the long-time limit after the voltage quench. The circles represent the result obtained by solving the time-local EOMs, while the solid line represents the result obtained from the Floquet Green's function method.
}
\label{Fig_Floquet} 
\end{figure}

Figures~\ref{Fig_ACJE}(a) and \ref{Fig_ACJE}(b) show that, after a sufficiently long time following the voltage quench, the current $J_1(t)$ relaxes to a periodic oscillation whose fundamental frequency is the conventional Josephson frequency $\omega_J=2eV$. This indicates that the system reaches the ac Josephson periodic steady state after passing through the nonequilibrium transient regime induced by the voltage quench. The ac Josephson steady state obtained in this long-time limit can also be calculated independently by the Floquet Green's function method~\cite{Sun2002}, which assumes a periodic steady state. In Fig.~\ref{Fig_Floquet}, we compare one period of $J_1(t)$ extracted from the long-time region of the time-local EOM calculation with the result obtained from the Floquet Green's function method described in Appendix~\ref{app.floquet}. The two results agree well, confirming that the time-local EOMs derived in this paper correctly describe the relaxation toward the ac Josephson steady state after the voltage quench.

\begin{figure}[t]
\centering
\includegraphics[width=\linewidth]{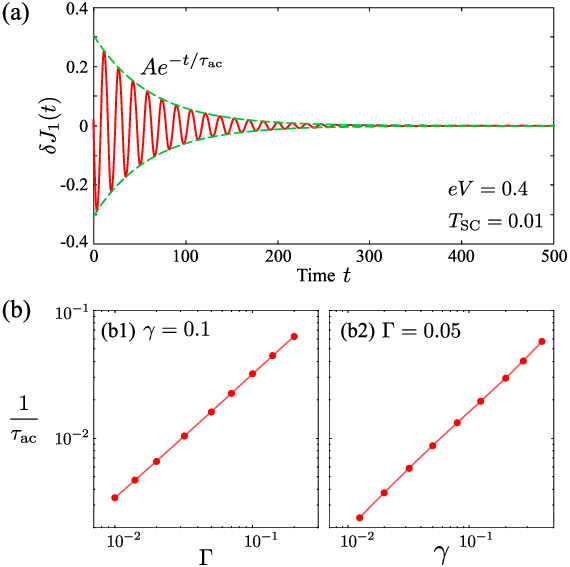}
\caption{
Relaxation toward the ac Josephson steady state. (a) Time evolution of $\delta J_1(t)$ for $eV=0.4$, $T_{\rm SC}=0.01$, $\Gamma=0.05$, and $\gamma=0.1$. The green dashed curves show the fitted exponential envelope $\pm A e^{-t/\tau_{\rm ac}}$, from which the relaxation time $\tau_{\rm ac}$ is extracted. (b) Relaxation rate $1/\tau_{\rm ac}$ as a function of (b1) the Dynes parameter $\Gamma$ at fixed $\gamma=0.1$ and (b2) the tunnel coupling $\gamma$ at fixed $\Gamma=0.05$. The other parameters are the same as in panel (a)
}
\label{Fig_tau}
\end{figure}

\begin{figure}[t]
\centering
\includegraphics[width=\linewidth]{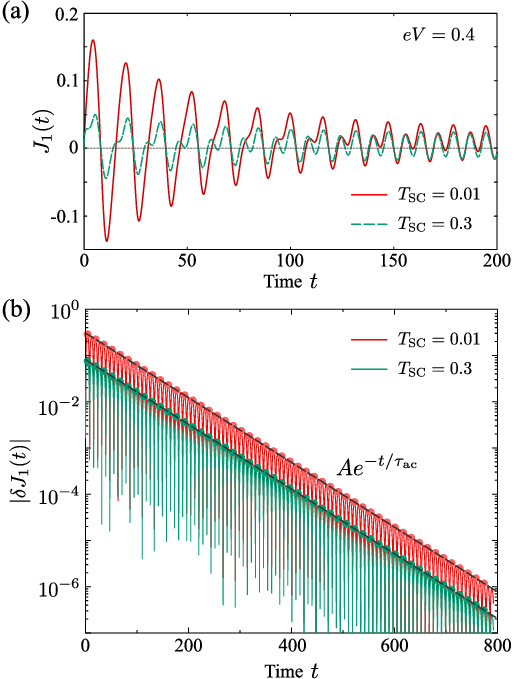}
\caption{
Temperature dependence of the voltage-quench dynamics for $eV=0.4$. (a) Comparison of the time evolution of the current $J_1(t)$ for $T_{\rm SC}=0.01$ (solid line) and $T_{\rm SC}=0.3$ (dashed line). (b) Decay of the transient-current component $|\delta J_1(t)|$ at the two temperatures shown in panel (a). The circles denote the successive maxima of $|\delta J_1(t)|$, and the straight dashed lines show fits to the exponential envelope $A e^{-t/\tau_{\rm ac}}$. The nearly parallel envelopes show that the relaxation time is almost independent of $T_{\rm SC}$, whereas the transient amplitude is reduced at the higher temperature.
}
\label{Fig_T}
\end{figure}

An important advantage of the present method is that it provides direct access to the transient dynamics before the system relaxes to the periodic steady state, which is not captured by the Floquet Green's function method. Indeed, Figs.~\ref{Fig_ACJE}(a) and \ref{Fig_ACJE}(b) show that the current oscillation immediately after the quench is dominated not by the Josephson frequency $\omega_J$ but by its subharmonic component $\omega_J/2$. Such a fractional Josephson oscillation appearing immediately after a voltage quench has recently been proposed as the nonequilibrium fractional Josephson effect (NFJE)~\cite{Lahiri2023}. In contrast to the conventional fractional Josephson effect originating from a $4\pi$-periodic current–phase relation in equilibrium~\cite{Kitaev2001, FuKane2009, Alicea2012, Beenakker2013}, the NFJE originates from nonequilibrium intermediate quasiparticle processes generated by a sharp voltage change. Microscopically, Cooper-pair tunneling can be viewed as a process in which two intermediate quasiparticles are virtually excited, tunnel across the junction, and subsequently recombine into a Cooper pair. After the voltage quench, intermediate quasiparticle processes occurring at different times can interfere with each other through the voltage-induced phase in the tunneling term and the retardation encoded in the memory kernel of the superconducting leads. As a result, the pair-current amplitude can be modulated at around $\omega_J/2$, giving rise to a transient fractional Josephson oscillation. The $\omega_J/2$ component observed immediately after the bias quench in Fig.~\ref{Fig_ACJE} can thus be interpreted as an NFJE-like transient response.

To quantify relaxation toward the ac Josephson steady state, we exploit the periodicity of the steady-state current, $J_1(t+T_J)=J_1(t)$, where $T_J=2\pi/\omega_J$ is the Josephson period. We define
\begin{equation}
\delta J_1(t)\equiv J_1(t+T_J)-J_1(t).
\end{equation}
This difference removes the periodic steady-state component from the real-time current $J_1(t)$ and thus isolates its transient part. As shown in Fig.~\ref{Fig_tau}(a), the envelope of $|\delta J_1(t)|$ is well described by
\begin{equation}
|\delta J_1(t)|_{\rm env}=A e^{-t/\tau_{\rm ac}}.
\end{equation}
We extract the relaxation time $\tau_{\rm ac}$ by fitting the successive local maxima of $|\delta J_1(t)|$ to this exponential form.

Panels (b1) and (b2) of Fig.~\ref{Fig_tau} show the relaxation rate $1/\tau_{\rm ac}$ as a function of the Dynes parameter $\Gamma$ and the tunnel coupling $\gamma$, respectively. In the subgap regime, direct quasiparticle escape into the continuum ($|\omega|>\Delta$) is strongly suppressed. The deviation from the periodic steady state thus decays mainly through coupling to the subgap states of the superconducting leads that are induced by the Dynes broadening. The parameter $\Gamma$ determines the damping of these lead states, while $\gamma$ controls their hybridization with the QD. Increasing either parameter thus accelerates the decay of the transient component and enhances $1/\tau_{\rm ac}$.

So far, we have fixed the lead temperature at $T_{\rm SC}=0.01$, but the present scheme also allows us to vary the temperature of the superconducting leads. In Fig.~\ref{Fig_T} (a), we compare the current responses for $T_{\rm SC}=0.01$ and $T_{\rm SC}=0.3$ at a fixed bias voltage $eV=0.4$. At both temperatures, an NFJE-like transient response appears immediately after the quench, although its amplitude is substantially smaller at the higher temperature. To distinguish this amplitude reduction from a change in the relaxation time, we plot $|\delta J_1(t)|$ together with the fitted exponential envelopes in Fig.~\ref{Fig_T}(b). The two envelopes have nearly the same slope on the logarithmic scale, indicating that $\tau_{\rm ac}$ is almost independent of $T_{\rm SC}$, whereas their different magnitudes reflect the temperature dependence of the transient amplitude. This behavior is consistent with the fact that the retarded lead self-energy, which governs the relaxation, does not depend on the Fermi distribution, whereas the temperature enters through the lesser self-energy and thus modifies only the weight of the transient response.

\begin{figure}[t]
\centering
\includegraphics[width=\linewidth]{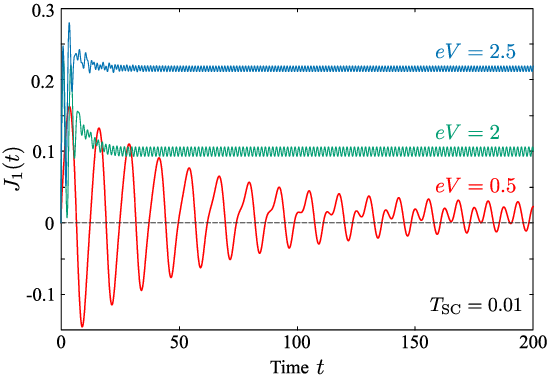}
\caption{
Bias-voltage dependence of the voltage-quench dynamics at $T_{\rm SC}=0.01$. The time evolution of the current $J_1(t)$ is shown for $eV=0.5$, $2$, and $2.5$.
}
\label{Fig_V} 
\end{figure}

We now turn to the bias-voltage dependence of the transient dynamics. The present scheme has no intrinsic restriction on the bias voltage, provided that the relevant energy window lies within the finite bandwidth $D$. In Fig.~\ref{Fig_V}, we compare the current responses for $eV=0.5$, $2$, and $2.5$ at a fixed temperature $T_{\rm SC}=0.01$. Since we set $\Delta=1$, $eV=2$ corresponds to the threshold at which quasiparticle excitation channels into the continuum begin to open. As shown in Fig.~\ref{Fig_V}, the transient dynamics after the quench differs significantly between the subgap regime, $eV<2\Delta$, and the above-gap regime, $eV\ge 2\Delta$. In the subgap regime, the subharmonic component near $\omega_J/2$ that appears immediately after the quench persists for a finite time, and the NFJE-like oscillation is clearly observed. In the above-gap regime, by contrast, direct quasiparticle escape into the continuum becomes energetically allowed, providing an additional and efficient relaxation channel for the transient subgap excitations. As a result, the $\omega_J/2$ component is strongly suppressed, and the current relaxes more rapidly to the ac Josephson steady state.

As demonstrated above, the present scheme properly describes the voltage-quench dynamics in an S-QD-S junction. In the long-time limit, the result agrees with the ac Josephson periodic steady state obtained from the Floquet Green's function method, reproducing the steady ac Josephson response whose fundamental frequency is the conventional Josephson frequency $\omega_J$. At the same time, because the present scheme directly solves the real-time evolution, it also describes the transient regime immediately after the quench. In the subgap-bias regime, we explicitly tracked the emergence of an NFJE-like transient oscillation with a subharmonic component near $\omega_J/2$, followed by relaxation to the conventional ac Josephson steady state. Although the discussion in this section has been restricted to a minimal model of a single-level, symmetric S-QD-S junction, the applicability of the present scheme is not limited to this setup. The central system may contain multiple orbitals or spatial degrees of freedom. The scheme can also be applied to more general time-dependent superconducting transport problems, including general voltage protocols $V_\alpha(t)$, time-dependent gate voltages or dot-level modulations, and multi-terminal Josephson junctions such as three-terminal and four-terminal setups.

\section{Summary}

In this work, we have derived time-local equations of motion within the nonequilibrium Green's function formalism for time-dependent quantum transport in systems coupled to superconducting leads. The self-energy of a superconducting lead is a strongly frequency-dependent Nambu matrix, reflecting both the gapped quasiparticle spectrum and anomalous correlations that mix particle and hole degrees of freedom. In the time domain, this frequency dependence gives rise to a nonlocal memory kernel, making direct real-time propagation of the Kadanoff–Baym equations computationally demanding. We addressed this problem by recasting the memory kernel into a time-local form while retaining the memory effects characteristic of superconducting leads.

To this end, we extended the auxiliary-mode expansion to Nambu-space self-energies for systems coupled to superconducting leads. Specifically, by combining the Pad\'e decomposition of the Fermi distribution with rational approximations based on the adaptive Antoulas–Anderson algorithm, we expressed the superconducting lead self-energy as a finite sum of exponential modes in the time domain. This expansion allowed us to transform the time-nonlocal convolution integrals appearing in the Kadanoff–Baym equations into a closed set of ordinary differential equations for the equal-time lesser Green's function, the current matrices, and the auxiliary matrices. The resulting equations are applicable to general time-dependent bias voltages, superconducting phases, and one-body Hamiltonians of the central system.

As a concrete application of the formulation developed here, we analyzed voltage-quench dynamics in an S-QD-S junction, where a single-level quantum dot is coupled to two superconducting leads. We showed that, after a dc bias is suddenly applied at $t=0$, the system evolves through a transient regime and eventually relaxes to a periodic steady state characterized by the ac Josephson frequency. This long-time behavior agrees with the periodic steady-state solution obtained from the Floquet Green's function method, confirming that the present approach correctly describes the real-time relaxation toward the ac Josephson steady state. The calculation also demonstrates that the transient dynamics before reaching the periodic steady state is directly accessible within the same framework.

The present scheme does not assume that the final state is known in advance or that the external driving is strictly periodic. It is therefore complementary to steady-state Floquet approaches and transient schemes that use the final periodic state as a reference, and can be applied to more general time-dependent driving protocols, including pulses, ramps, chirps, and noisy biases. Since the Hamiltonian of the central system can be treated as a general time-dependent one-body Hamiltonian, the scheme is not limited to quantum-dot systems but can be applied to a wide range of central systems, including semiconductor nanowires, two-dimensional electron systems, and mesoscopic superconducting systems. We expect that this formulation will provide a useful basis for studying a variety of nonequilibrium phenomena in superconducting hybrid structures, including Andreev-bound-state dynamics under nonperiodic driving, multifrequency Josephson responses in systems with multiple superconducting terminals, and self-consistent order-parameter dynamics in driven superconducting systems.

\begin{acknowledgments}
We thank Shuntaro Sumita and Yusuke Kato for stimulating discussions. T.K. was supported by MEXT and JSPS KAKENHI Grant-in-Aid for JSPS fellows Grant No.~25K23363.
\end{acknowledgments}

\section*{Data Availability}
The data that support the findings of this article are not publicly available. The data are available from the authors upon reasonable request.

\appendix
\section{Derivation of Eq.~\eqref{eq.EOM.GL}}
\label{sec.app.KB}

In this Appendix, we derive the equation of motion for the equal-time lesser Green's function, Eq.~\eqref{eq.EOM.GL}, starting from the Keldysh equation. The lesser and greater Green’s functions are written as~\cite{Datta1995, Haug2008, Stefanucci2013}
\begin{equation}
\scalebox{0.97}{$\displaystyle
\bm{G}^\lessgtr(t, t') =
\int_{-\infty}^\infty dt_1
\int_{-\infty}^\infty dt_2\,
\bm{G}^{\rm R}(t, t_1)
\bm{\Sigma}^\lessgtr(t_1, t_2)
\bm{G}^{\rm A}(t_2, t')$},
\label{eq.Keldysh.eq}
\end{equation}
where $\bm{G}^{\rm R}$ and $\bm{G}^{\rm A}$ are the retarded and advanced Green's functions of the central system dressed by the coupling to the superconducting leads. These Green's functions satisfy the Dyson equation in the form
\begin{align}
\bm{G}^{\rm R(A)}(t,t')
&=
\bm{G}^{\rm R(A)}_0(t,t')
+
\int_{-\infty}^\infty dt_1
\int_{-\infty}^\infty dt_2
\notag\\[4pt]
&\quad \times
\bm{G}^{\rm R(A)}_0(t,t_1)
\bm{\Sigma}^{\rm R(A)}(t_1,t_2)
\bm{G}^{\rm R(A)}(t_2,t'),
\label{eq.Dyson.right}	
\end{align}
or equivalently
\begin{align}
\bm{G}^{\rm R(A)}(t,t')
&=
\bm{G}^{\rm R(A)}_0(t,t')
+
\int_{-\infty}^\infty dt_1
\int_{-\infty}^\infty dt_2
\notag\\[4pt]
&\quad \times
\bm{G}^{\rm R(A)}(t,t_1)
\bm{\Sigma}^{\rm R(A)}(t_1,t_2)
\bm{G}^{\rm R(A)}_0(t_2,t').
\label{eq.Dyson.left}	
\end{align}
Here, $\bm{G}_0^{\rm R(A)}$ denotes the retarded (advanced) Green's function of the isolated central system, defined in the absence of the system–lead coupling.

To derive Eq.~\eqref{eq.EOM.GL} from the Keldysh–Dyson equations, it is useful to introduce the inverse operators of the bare Green's functions. We define the right- and left-acting inverse Green's functions,  $\overrightarrow{\bm{G}}_0^{-1}(t)$ and $\overleftarrow{\bm{G}}_0^{-1}(t)$, by
\begin{subequations}
\begin{align}
& 
\overrightarrow{\bm{G}}_0^{-1}(t) \bm{G}^{\rm R(A)}_0(t,t') =
\delta(t-t') \bm{I}_{2N}
,\\[4pt]
&
\bm{G}^{\rm R(A)}_0(t,t') \overleftarrow{\bm{G}}_0^{-1}(t') =
\delta(t-t') \bm{I}_{2N},
\end{align}
\end{subequations}
where $\bm{I}_{2N}$ is the $2N\times 2N$ identity matrix. These inverse operators are explicitly given by~\cite{Kawamura2024}
\begin{subequations}
\begin{align}
& 
\overrightarrow{\bm{G}}_0^{-1}(t) = 
i\overrightarrow{\frac{\partial}{\partial t}}\, \bm{I}_{2N}
-\bm{h}_{\rm sys}(t)
,\\[4pt]
&
\overleftarrow{\bm{G}}_0^{-1}(t') = 
-i\overleftarrow{\frac{\partial}{\partial t'}}\, \bm{I}_{2N}
-\bm{h}_{\rm sys}(t').
\end{align}
\end{subequations}
Here, the arrows indicate the direction in which the time derivatives act.

Acting with $\overrightarrow{\bm{G}}_0^{-1}(t)$ on Eq.~\eqref{eq.Dyson.right} and with $\overleftarrow{\bm{G}}_0^{-1}(t')$ on Eq.~\eqref{eq.Dyson.left}, respectively, we obtain
\begin{subequations}
\begin{align}
\overrightarrow{\bm{G}}_0^{-1}(t)\, 
\bm{G}^{\rm R}(t,t') 
&=
\delta(t-t')\, \bm{I}_{2N} 
\notag\\[4pt]
&\quad +
\int_{-\infty}^\infty dt_1\,
\bm{\Sigma}^{\rm R}(t, t_1) \bm{G}^{\rm R}(t_1, t')
\label{eq.G0GR}
,\\
\bm{G}^{\rm A}(t, t')\,
\overleftarrow{\bm{G}}_0^{-1}(t') 
&= 
\delta(t-t')\, \bm{I}_{2N} 
\notag\\[4pt]
&\quad +
\int_{-\infty}^\infty dt_1\,
\bm{G}^{\rm A}(t, t_1)
\bm{\Sigma}^{\rm A}(t_1, t').
\label{eq.GAG0}
\end{align}
\end{subequations}
In the same manner, acting with the inverse operators on the Keldysh equation~\eqref{eq.Keldysh.eq} and using Eqs.~\eqref{eq.G0GR} and \eqref{eq.GAG0}, we obtain the Kadanoff–Baym equations~\cite{Datta1995, Haug2008, Stefanucci2013}
\begin{subequations}
\begin{align}
& 
i\frac{\partial}{\partial t}
\bm{G}^\lessgtr(t,t') =
\bm{h}_{\rm sys}(t)\,
\bm{G}^\lessgtr(t,t') 
\notag\\[4pt]
&\quad
+
\int_{-\infty}^\infty dt_1\,
\Big[
\bm{\Sigma}^{\rm R}(t, t_1) \bm{G}^\lessgtr(t_1, t')  +
\bm{\Sigma}^\lessgtr(t, t_1) \bm{G}^{\rm A}(t_1, t')
\Big]
\label{eq.GL.left}
,\\
&
i\frac{\partial}{\partial t'}
\bm{G}^\lessgtr(t, t')\,
=
-\bm{G}^\lessgtr(t,t')\,
\bm{h}_{\rm sys}(t')
\notag\\[4pt]
&\quad -
\int_{-\infty}^\infty dt_1\,
\Big[
\bm{G}^{\rm R}(t, t_1) \bm{\Sigma}^\lessgtr(t_1, t') +
\bm{G}^\lessgtr(t, t_1) \bm{\Sigma}^{\rm A}(t_1, t') 
\Big].
\label{eq.GL.right}
\end{align}
\end{subequations}
The retarded and advanced components are related to the lesser and greater components as
\begin{subequations}
\begin{align}
&
\bm{X}^{\rm R}(t,t') = 
\Theta(t-t') \big[\bm{X}^>(t,t') -\bm{X}^<(t,t') \big]
,\\[4pt]
&
\bm{X}^{\rm A}(t,t') = 
-\Theta(-t +t') \big[\bm{X}^>(t,t') -\bm{X}^<(t,t') \big],
\end{align}
\end{subequations}
where $\bm{X}=\bm{G}$ or $\bm{\Sigma}$. Using these relations and taking the equal-time limit $t'=t$, the collision integrals in Eqs.~\eqref{eq.GL.left} and \eqref{eq.GL.right} can be expressed in terms of the current matrix $\bm{\Pi}_\alpha(t)$ defined in Eq~\eqref{eq.Pi}. We thus find
\begin{subequations}
\begin{align}
i\left.
\frac{\partial}{\partial t}
\bm{G}^<(t,t')
\right|_{t'=t}
&=
\bm{h}_{\rm sys}(t)\,
\bm{G}^<(t,t)
-
\sum_{\alpha=1}^{N_{\rm lead}}
\bm{\Pi}^\dagger_\alpha(t)
\label{eq.GL.left2}
,\\
i\left.
\frac{\partial}{\partial t'}
\bm{G}^<(t,t')
\right|_{t'=t}
&=
-\bm{G}^<(t,t)\,
\bm{h}_{\rm sys}(t)
-
\sum_{\alpha=1}^{N_{\rm lead}}
\bm{\Pi}_\alpha(t).
\label{eq.GL.right2}
\end{align}
\end{subequations}
Adding these two equations yields Eq.~\eqref{eq.EOM.GL}.

\section{Derivation of Eq.~\eqref{eq.dt.Pi}}
\label{sec.app.dt.Pi}

In this Appendix, we derive Eq.~\eqref{eq.dt.Pi}. Taking the time derivative of $\bm{\Pi}_{\alpha,\nu,l}(t)$ defined in Eq.~\eqref{eq.Pi.AME}, we obtain
\begin{align}
i \frac{d}{dt}\, \bm{\Pi}_{\alpha, \nu, l}(t)
&=
i \frac{d}{dt}
\int_{-\infty}^t dt_1\,
\big[
\bm{G}^>(t,t_1) \bm{\Sigma}^<_{\alpha,\nu, l}(t_1,t) 
\notag\\
&\hspace{2.1cm}
-
\bm{G}^<(t,t_1) \bm{\Sigma}^>_{\alpha,\nu, l}(t_1,t)
\big]
\notag\\[6pt]
&=
i\big[
\bm{G}^>(t,t^-) \bm{\Sigma}^<_{\alpha,\nu, l}(t^-,t) 
\notag\\
&\hspace{2.1cm}
-
\bm{G}^<(t,t^-) \bm{\Sigma}^>_{\alpha,\nu, l}(t^-,t)
\big]
\notag\\
&\qquad
+
\int_{-\infty}^t dt_1\,
\big[
i\partial_t\bm{G}^>(t,t_1)\, 
\bm{\Sigma}^<_{\alpha,\nu, l}(t_1,t) 
\notag\\
&\hspace{2.1cm}
-
i\partial_t\bm{G}^<(t,t_1)\, 
\bm{\Sigma}^>_{\alpha,\nu, l}(t_1,t)
\big]
\notag\\
&\qquad
+
\int_{-\infty}^t dt_1\,
\big[
\bm{G}^>(t,t_1)\, 
i\partial_t \bm{\Sigma}^<_{\alpha,\nu, l}(t_1,t) 
\notag\\
&\hspace{2.1cm}
-
\bm{G}^<(t,t_1)\, 
i\partial_t \bm{\Sigma}^>_{\alpha,\nu, l}(t_1,t)
\big].
\label{eq.dt.Pi.1}
\end{align}
Here, $t^-$ denotes a time infinitesimally earlier than $t$. The derivative $i\partial_t\, \bm{G}^\lessgtr(t,t_1)$ is evaluated using the Kadanoff–Baym equation~\eqref{eq.GL.left}. For $t_1<t$, the derivative of the mode-resolved self-energy is obtained from Eqs.~\eqref{eq.Kernel.AME.def} and \eqref{eq.self.AME.2} as
\begin{align}
i\frac{\partial}{\partial t}
\bm{\Sigma}^\lessgtr_{\alpha, \nu, l}(t_1, t)
&=
\scalebox{0.95}{$\displaystyle
\bigg[
i\frac{\partial}{\partial t}
\m{K}^\lessgtr_{\alpha,\nu,l}(t_1 -t)
\bigg]
\ket{{u}_{\alpha,\nu}(t_1)}
\bra{{u}_{\alpha,\nu}(t)}
$}
\notag\\
&\quad
\scalebox{0.95}{$\displaystyle
+
\m{K}^\lessgtr_{\alpha,\nu,l}(t_1 -t)
\ket{{u}_{\alpha,\nu}(t_1)}
\bigg[
i\frac{\partial}{\partial t}
\bra{{u}_{\alpha,\nu}(t)}
\bigg]$}
\notag\\[4pt]
&=
\scalebox{0.95}{$\displaystyle
-\chi_{\alpha,\nu,l}\, \bm{\Sigma}^\lessgtr_{\alpha, \nu, l}(t_1, t)
-\bm{\Sigma}^\lessgtr_{\alpha, \nu, l}(t_1, t)\, \bm{\Phi}_\alpha(t)
$}.
\label{eq.EOM.self.lg}
\end{align}
Substituting Eqs.~\eqref{eq.GL.left} and \eqref{eq.EOM.self.lg} into Eq.~\eqref{eq.dt.Pi.1}, we find
\begin{align}
i\frac{d}{dt}\, \bm{\Pi}_{\alpha,\nu,l}(t)
&=
i\big[
\bm{G}^>(t,t^-) \bm{\Sigma}^<_{\alpha,\nu, l}(t^-,t) 
\notag\\
&\hspace{1cm}
-
\bm{G}^<(t,t^-) \bm{\Sigma}^>_{\alpha,\nu, l}(t^-,t)
\big]
\notag\\
&\quad
+
\big[\bm{h}_{\rm sys}(t)- \chi_{\alpha,\nu,l}\big]
\bm{\Pi}_{\alpha,\nu,l}(t)
\notag\\
&\quad
-
\bm{\Pi}_{\alpha,\nu,l}(t)\,
\bm{\Phi}_\alpha(t)
+
\bm{\Omega}_{\alpha,\nu,l}(t),
\label{eq.dt.Pi.2}
\end{align}
where the auxiliary matrix $\bm{\Omega}_{\alpha,\nu,l}(t)$ is defined in Eq.~\eqref{eq.def.Omega}.

It remains to evaluate the boundary term in Eq.~\eqref{eq.dt.Pi.2}. Since $t^-<t$, Eqs.~\eqref{eq.Kernel.AME.def} and \eqref{eq.self.AME.2} give
\begin{align}
&
i
\big[
\bm{G}^>(t,t^-) \bm{\Sigma}^<_{\alpha,\nu, l}(t^-,t)
-
\bm{G}^<(t,t^-) \bm{\Sigma}^>_{\alpha,\nu, l}(t^-,t)
\big]
\notag\\[2pt]
=&
-R^<_{\alpha,\nu,l}
\ket{{u}_{\alpha,\nu}(t)}
\bra{{u}_{\alpha,\nu}(t)}
\notag\\
&\quad
+
i\big[R^>_{\alpha,\nu,l} - R^<_{\alpha,\nu,l}\big]
\bm{G}^<(t) 
\ket{{u}_{\alpha,\nu}(t)}
\bra{{u}_{\alpha,\nu}(t)}.
\label{eq.dt.Pi.3}
\end{align}
Here, we have used the equal-time relation
\begin{equation}
\bm{G}^>(t,t^-)
=
-i\bm{I}_{2N}+
\bm{G}^<(t).
\label{eq.equal.time.relation}
\end{equation}
Substituting Eq.~\eqref{eq.dt.Pi.3} into Eq.~\eqref{eq.dt.Pi.2} and using the definition of $\bm{P}_{\alpha,\nu}(t)$ in Eq.~\eqref{eq.def.P}, we arrive at Eq.~\eqref{eq.dt.Pi}.

\section{Derivation of the equation of motion for $\bm{\Omega}_{K,K'}(t)$}
\label{sec.app.EOM.Omega}

In this Appendix, we derive the equation of motion for the auxiliary matrix $\bm{\Omega}_{K,K'}(t)$. Taking the time derivative of $\bm{\Omega}_{K,K'}(t)$ defined in Eq.~\eqref{eq.Omega.AME}, we obtain three contributions,
\begin{equation}
i \frac{d}{dt}\, \bm{\Omega}_{K, K'}(t)=
\bm{\m{I}}_1(t) + \bm{\m{I}}_2(t) +\bm{\m{I}}_3(t).
\end{equation}
The first term, $\bm{\m{I}}_1(t)$, originates from the time derivative of the upper limits of the two time integrals. The second term, $\bm{\m{I}}_2(t)$, arises from the derivative of the self-energy whose first time argument is $t$, namely $\partial_t\,\bm{\Sigma}^{\lessgtr}_{K'}(t,t_2)$. The third term, $\bm{\m{I}}_3(t)$, arises from the derivative of the self-energy whose second time argument is $t$, namely $\partial_t\, \bm{\Sigma}^{\lessgtr}_{K}(t_1,t)$.

The boundary contribution $\bm{\m{I}}_1(t)$ can be evaluated using the equal-time relation in Eq.~\eqref{eq.equal.time.relation} and the definition of the mode-resolved current matrix in Eq.~\eqref{eq.Pi.AME}. We obtain
\begin{widetext}
\begin{align}
\bm{\m{I}}_1(t)
&=
i\int_{-\infty}^t dt_2\,
\Big[
\bm{\Sigma}^>_{K'}(t, t_2)
-
\bm{\Sigma}^<_{K'}(t, t_2)
\Big]
\Big[
\bm{G}^>(t_2,t)
\bm{\Sigma}^<_{K}(t^-,t)
-
\bm{G}^<(t_2,t)\bm{\Sigma}^>_{K}(t^-,t)
\Big]
\notag\\
&\quad
-
i\int_{-\infty}^{t} dt_2\,
\Big[
\bm{\Sigma}^<_{K'}(t, t_2)
\bm{G}^{\rm A}(t_2,t)
\bm{\Sigma}^>_{K}(t^-,t)
-
\bm{\Sigma}^>_{K'}(t, t_2)
\bm{G}^{\rm A}(t_2,t)
\bm{\Sigma}^<_{K}(t^-,t)
\Big]
\notag\\
&\qquad
+
i\Big[
\bm{\Sigma}^>_{K'}(t, t^-)
-
\bm{\Sigma}^<_{K'}(t, t^-)
\Big]
\int_{-\infty}^t dt_1\,
\Big[
\bm{G}^>(t,t_1)
\bm{\Sigma}^<_{K}(t_1,t)
-
\bm{G}^<(t,t_1)
\bm{\Sigma}^>_{K}(t_1,t)
\Big]
\notag\\[8pt]
&=
i\int_{-\infty}^t dt_2\,
\Big[
\bm{\Sigma}^>_{K'}(t, t_2)-
\bm{\Sigma}^<_{K'}(t, t_2)
\Big]
\Big[
\bm{G}^>(t_2,t)
\bm{\Sigma}^<_{K}(t^-,t)
-
\bm{G}^<(t_2,t)\bm{\Sigma}^>_{K}(t^-,t)
\Big]
\notag\\
&\quad
+
i\int_{-\infty}^{t} dt_2\,
\Big[
\bm{\Sigma}^<_{K'}(t, t_2)
\big[
\bm{G}^>(t_2,t) -\bm{G}^<(t_2,t)
\big]
\bm{\Sigma}^>_{K}(t^-,t)
-
\bm{\Sigma}^>_{K'}(t, t_2)
\big[
\bm{G}^>(t_2,t) -\bm{G}^<(t_2,t)
\big]
\bm{\Sigma}^<_{K}(t^-,t)
\Big]
\notag\\
&\qquad
+
i\Big[
\bm{\Sigma}^>_{K'}(t, t^-)
-
\bm{\Sigma}^<_{K'}(t, t^-)
\Big]
\bm{\Pi}_K(t)
\notag\\[8pt]
=&
-i \int_{-\infty}^t dt_1\,
\Big[
\bm{\Sigma}^<_{K'}(t, t_1) \bm{G}^>(t_1, t) -
\bm{\Sigma}^>_{K'}(t, t_1) \bm{G}^<(t_1, t)
\Big]\,
\bm{\Sigma}^<_{K}(t^-,t)
\notag\\
&\quad
+i \int_{-\infty}^t dt_1\,
\Big[
\bm{\Sigma}^<_{K'}(t, t_1) \bm{G}^>(t, t_1) -
\bm{\Sigma}^>_{K'}(t, t_1) \bm{G}^<(t, t_1)
\Big](t,t)\,
\bm{\Sigma}^>_{K}(t^-,t)
+
i\Big[
\bm{\Sigma}^>_{K'}(t, t^-)-
\bm{\Sigma}^<_{K'}(t, t^-)
\Big]
\bm{\Pi}_{K}(t)
\notag\\[8pt]
=&
-i
\big[R^>_{K} -R^<_{K}\big]\, 
\bm{\Pi}^\dagger_{K'}(t)\,
\bm{P}_{K}(t)
+
i
\big[\big(R^>_{K'}\big)^* -\big(R^<_{K'}\big)^*\big]
\bm{P}_{K'}(t)\,
\bm{\Pi}_{K}(t).
\label{eq.I1.Omega}
\end{align}
\end{widetext}
This contribution gives the inhomogeneous term in Eq.~\eqref{eq.EOM.Omega}.

Next, we evaluate $\bm{\m{I}}_2(t)$. For $t>t_2$, the time derivative of the mode-resolved self-energy is given by
\begin{equation}
i\frac{\partial}{\partial t}\,
\bm{\Sigma}^\lessgtr_{K'}(t,t_2)
=
\chi^*_{K'} \bm{\Sigma}^\lessgtr_{K'}(t,t_2)
+
\bm{\Phi}_{K'}(t)\,
\bm{\Sigma}^\lessgtr_{K'}(t,t_2).
\end{equation}
Using this relation, we find
\begin{equation}
\bm{\m{I}}_2(t) =
\chi^*_{K'}\,\bm{\Omega}_{K,K'}(t)+
\bm{\Phi}_{K'}(t)\,\bm{\Omega}_{K,K'}(t).
\label{eq.I2.Omega}
\end{equation}

Similarly, $\bm{\m{I}}_3(t)$ is evaluated by using Eq.~\eqref{eq.EOM.self.lg}, which describes the time derivative of the self-energy with respect to its second time argument. This gives
\begin{equation}
\bm{\m{I}}_3(t) =
-\chi_{K}\,\bm{\Omega}_{K,K'}(t)
-\bm{\Omega}_{K,K'}(t)\, \bm{\Phi}_{K}(t).
\label{eq.I3.Omega}
\end{equation}
Combining the three contributions in Eqs.~\eqref{eq.I1.Omega}, \eqref{eq.I2.Omega}, and \eqref{eq.I3.Omega}, we obtain Eq.~\eqref{eq.EOM.Omega}.

\section{Floquet Green's function method for the S-QD-S junction}
\label{app.floquet}

In this Appendix, we summarize the Floquet Green's function method~\cite{Sun2002} used to benchmark the long-time dynamics obtained from the time-local EOMs~\eqref{eq.EOM.GL}–\eqref{eq.EOM.Omega}. We focus on the minimal S-QD-S junction considered in Sec.~\ref{sec.application}.

In the long-time limit after a dc voltage bias $V$ is applied, the system reaches a periodic ac Josephson state. In this periodic steady state, the dot Green's functions and the self-energies due to the superconducting leads satisfy
\begin{subequations}
\begin{align}
&
\bm{G}^{X}(t+T,\, t'+T)=
\bm{G}^{X}(t,t')
,\\[4pt]
&
\bm{\Sigma}^{X}(t+T,\, t'+T)=
\bm{\Sigma}^{X}(t,t'),
\end{align}
\end{subequations}
where $X={\rm R},{\rm A},<$. The fundamental frequency used in the Floquet representation is
\begin{equation}
\Omega=eV=\frac{\omega_J}{2},
\end{equation}
and the corresponding period is
\begin{equation}
T = \frac{2\pi}{\Omega}.
\end{equation}
Here, $\omega_J=2eV$ is the Josephson frequency. The central idea of the Floquet Green's function method is to exploit this periodicity and rewrite the Keldysh equation~\eqref{eq.Keldysh.eq} and the Dyson equation~\eqref{eq.Dyson.right} in Floquet space as~\cite{Sun2002}
\begin{subequations}
\begin{align}
& 
\scalebox{0.97}{$\displaystyle
\bm{G}^<_{mn}(\omega) =
\sum_{k,l}
\bm{G}^{\rm R}_{ml}(\omega)
\bm{\Sigma}^<_{lk}(\omega)
\bm{G}^{\rm A}_{kn}(\omega)$}
\label{eq.Keldysh.FL}
,\\
&
\scalebox{0.97}{$\displaystyle
\bm{G}^{\rm R(A)}_{mn}(\omega) =
\bm{G}^{\rm R(A)}_{0,mn}(\omega) +
\sum_{k,l}
\bm{G}^{\rm R(A)}_{0,ml}(\omega)
\bm{\Sigma}^{\rm R(A)}_{lk}(\omega)
\bm{G}^{\rm R(A)}_{kn}(\omega).$}
\label{eq.Dyson.GR.Floquet}
\end{align}
\end{subequations}
Here, for any two-time quantity satisfying
$
\bm{A}(t+T,\, t'+T) =
\bm{A}(t,t'),
$
we define its Floquet representation by~\cite{Sun2002}
\begin{align}
\bm{A}_{mn}(\omega)
&=
\frac{1}{T}
\int_{-T/2}^{T/2} dt_{\rm av}
\int_{-\infty}^{\infty} dt_{\rm rel}
\notag\\[4pt]
&\quad\times
e^{i(m-n)\Omega t_{\rm av}}\,
e^{i[\omega+(m+n)\Omega/2]t_{\rm rel}}
\bm{A}(t,t'),
\label{eq.floquet.def}	
\end{align}
with
\begin{equation}
t_{\rm av}=\frac{t+t'}{2},
\quad
t_{\rm rel}=t-t'.
\end{equation}
In this representation, the quasienergy $\omega$ is restricted to the first Floquet Brillouin zone,
\begin{equation}
-\frac{\Omega}{2}<\omega<\frac{\Omega}{2}.
\end{equation}

Using the definition of the Floquet transformation in Eq.~\eqref{eq.floquet.def}, the bare Green's function of the isolated dot is obtained as
\begin{equation}
\bm{G}^{\rm R(A)}_{0,mn}(\omega)=
\begin{pmatrix}
\dfrac{1}{\omega_m \pm i0^+ -\ep_{\rm d}} & 0 \\
0 & \dfrac{1}{\omega_m \pm i0^+ +\ep_{\rm d}}
\end{pmatrix}
\delta_{m,n},
\label{eq.Floquet.G0R}
\end{equation}
where $\omega_m \equiv \omega+m\Omega$ and $\ep_{\rm d}$ is the dot level.

The time-domain retarded and advanced self-energies have the same form as Eq.~\eqref{eq.self}. For the present single-level dot ($N=1$), they are written as
\begin{equation}
\bm{\Sigma}^{\rm R(A)}_\alpha(t, t')
=
|\m{T}_\alpha|^2
\sum_{\bm{k}}\,
U_\alpha(t)
\tau_z\,
D_{\alpha}^{\rm R(A)}(\bm{k},t -t')\,
\tau_z\,
U^\dagger_\alpha(t').
\label{eq.self.R}
\end{equation}
From Eqs.~\eqref{eq.U} and \eqref{eq.phi.application}, the phase matrix is given by
\begin{equation}
U_\alpha(t)=
\begin{pmatrix}
e^{-i \eta_\alpha \Omega t/2} & 0 \\
0 & e^{i \eta_\alpha \Omega t/2}
\end{pmatrix}.
\label{eq.floquet.U}
\end{equation}
Here, we have introduced the sign factor
\begin{equation}
\eta_\alpha
=
\begin{cases}
+1 & (\alpha=1),\\
-1 & (\alpha=2),
\end{cases}
\end{equation}
and set the initial phases to $\phi_1(0)=\phi_2(0)=0$, as in the voltage-quench protocol considered in Sec.~\ref{sec.application}. After performing the momentum summation in Eq.~\eqref{eq.self.R} and then applying the Floquet transformation, we obtain
\begin{align}
&
\bm{\Sigma}^{\rm R}_{\alpha, mn}(\omega)=
-i \gamma_\alpha
\notag\\
&\times
\scalebox{0.94}{$\displaystyle
\begin{pmatrix}
\beta_\alpha(\omega_{m -\eta_\alpha/2})\, \delta_{m,\, n} &
\dfrac{\beta_\alpha(\omega_{m -\eta_\alpha/2})\, \Delta_\alpha}{\omega_{m -\eta_\alpha/2} +i\Gamma_\alpha}\,
\delta_{m,n +\eta_\alpha}\\[8pt]
\dfrac{\beta_\alpha(\omega_{m +\eta_\alpha/2})\, \Delta_\alpha}{\omega_{m +\eta_\alpha/2} +i\Gamma_\alpha}\,
\delta_{m,n -\eta_\alpha}
&
\beta_\alpha(\omega_{m +\eta_\alpha/2})\, \delta_{m,\, n}
\end{pmatrix}$}
\label{eq.selfR.Floquet}
,\\
&
\bm{\Sigma}^{\rm A}_{\alpha, mn}(\omega)=
\big[
\bm{\Sigma}^{\rm R}_{\alpha, nm}(\omega)
\big]^\dagger.
\end{align}
Here, we have introduced the shorthand notation
\begin{equation}
\omega_{m \pm 1/2} \equiv
\omega+\left(m\pm\frac{1}{2}\right)\Omega.
\end{equation}
In the same manner, the lesser self-energy is obtained as
\begin{widetext}
\begin{equation}
\bm{\Sigma}^<_{\alpha, mn}(\omega) = 
2i\gamma_\alpha
\begin{pmatrix}
f_\alpha(\omega_{m-\eta_\alpha/2})\,
{\rm Re}\big[\beta_\alpha(\omega_{m-\eta_\alpha/2})\big]\,
\delta_{m,\, n}
&
f_\alpha(\omega_{m-\eta_\alpha/2})\,
{\rm Re}\left[\dfrac{\beta_\alpha(\omega_{m-\eta_\alpha/2})\, \Delta_\alpha}{\omega_{m-\eta_\alpha/2} +i\Gamma_\alpha}\right]
\delta_{m,n+\eta_\alpha}
\\[8pt]
f_\alpha(\omega_{m+\eta_\alpha/2})\,
{\rm Re}\left[\dfrac{\beta_\alpha(\omega_{m+\eta_\alpha/2})\, \Delta_\alpha}{\omega_{m+\eta_\alpha/2} +i\Gamma_\alpha}\right]
\delta_{m,n-\eta_\alpha}
&
f_\alpha(\omega_{m+\eta_\alpha/2})\,
{\rm Re}\big[\beta_\alpha(\omega_{m+\eta_\alpha/2})\big]\,
\delta_{m,\, n}
\end{pmatrix}.
\label{eq.selfL.Floquet}
\end{equation}
\end{widetext}
The total self-energies are obtained by summing over the two leads,
\begin{equation}
\bm{\Sigma}^{X={\rm R},{\rm A},<}_{mn}(\omega)
=
\sum_{\alpha=1,2}
\bm{\Sigma}^{X}_{\alpha,mn}(\omega).
\label{eq.selfL.Floquet.tot}
\end{equation}
Substituting Eqs.~\eqref{eq.Floquet.G0R}, \eqref{eq.selfR.Floquet}, \eqref{eq.selfL.Floquet}, and \eqref{eq.selfL.Floquet.tot} into the Keldysh equation~\eqref{eq.Keldysh.FL} and the Dyson equation~\eqref{eq.Dyson.GR.Floquet}, we obtain the dressed Floquet Green's functions $\bm{G}^{{\rm R, A},<}_{mn}(\omega)$.  In numerical calculations, the Floquet indices are truncated to a finite range, and convergence is checked by increasing the cutoff in Floquet space.

Once the dressed Floquet Green's functions are obtained, we evaluate the current in the ac Josephson steady state. In the Floquet representation, the current matrix in Eq.~\eqref{eq.Pi} is written as
\begin{equation}
\bm{\Pi}_{\alpha,mn}(\omega)=
\sum_{l}
\Big[
\bm{G}^{\rm R}_{ml}(\omega) \bm{\Sigma}^<_{\alpha,ln}(\omega) +
\bm{G}^<_{ml}(\omega) \bm{\Sigma}^{\rm A}_{\alpha,ln}(\omega) 
\Big].
\end{equation}
The time-dependent current matrix is then reconstructed as
\begin{equation}
\bm{\Pi}_\alpha(t)
=
\sum_{m,n}
e^{-i(m-n)\Omega t}
\int_{-\Omega/2}^{\Omega/2}
\frac{d\omega}{2\pi}\,
\bm{\Pi}_{\alpha,mn}(\omega).
\label{eq.floquet.Pi.time}
\end{equation}
Using Eq.~\eqref{eq.current.J}, the current is given by
\begin{equation}
J_\alpha(t) = 2e\, {\rm Re} 
\sum_{m,n} e^{-i(m-n)\Omega t}
\int_{-\Omega/2}^{\Omega/2} \frac{d\omega}{2\pi}\,
{\rm Tr}\big[\tau_z\, \bm{\Pi}_{\alpha,mn}(\omega)\big].
\end{equation}

\bibliography{AC_Josephson}
\end{document}